\documentclass[a4paper,12pt]{article}

\usepackage{amsmath}
\usepackage{amssymb}
\usepackage{graphicx}
\usepackage{hyperref}
\usepackage{cite}
\usepackage{color}

\makeatletter
\@addtoreset{equation}{section}
\renewcommand{\theequation}{\thesection.\@arabic\c@equation}
\makeatother

\begin{document}

\begin{titlepage}

\vspace*{-15mm}   
\baselineskip 10pt   
\begin{flushright}   
\begin{tabular}{r}    
\end{tabular}   
\end{flushright}   
\baselineskip 24pt   
\vglue 10mm   

\begin{center}
{\Large\bf
 Fluid model of a black hole--string transition
}

\vspace{8mm}   

\baselineskip 18pt   

\renewcommand{\thefootnote}{\fnsymbol{footnote}}

Yoshinori Matsuo\footnote[2]{ymatsuo@phys.kindai.ac.jp} 

\renewcommand{\thefootnote}{\arabic{footnote}}
 
\vspace{5mm}   

{\it  
 Department of Physics, Kindai University, \\Higashi-Osaka, Osaka 577-8502, Japan
}
  
\vspace{10mm}   

\end{center}

\begin{abstract}

A fluid model of self-gravitating strings is proposed. 
It is expected that black holes turn into strings 
around the end of black hole evaporation. 
The transition will occur near the Hagedorn temperature. 
After the transition, strings would form a bound state by the self-gravitation. 
Horowitz and Polchinski formulated a model of self-gravitating strings 
by using winding strings wrapping on the Euclidean time circle \cite{Horowitz:1997jc}. 
In this paper, we first show that winding strings in the Horowitz-Polchinski model 
approximately behave as a perfect fluid. 
Then, we solve the Einstein equation for the fluid of winding strings. 
Our solution reproduces behaviors of the self-gravitating string solution 
in the Horowitz-Polchinski model near the Hagedorn temperature, while 
it approaches the Schwarzschild black hole at low temperatures. 
Thus, our fluid model of self-gravitating strings gives a description 
of the transition between black holes and strings. 

\end{abstract}

\baselineskip 18pt   

\end{titlepage}

\newpage

\baselineskip 18pt

\tableofcontents


\section{Introduction}\label{sec:Intro}

Information loss paradox is one of the most important problems 
in black hole physics \cite{Hawking:1975vcx, Hawking:1976ra}. 
In order to solve this problem, it would be important 
to understand what happens in the last stages of black hole evaporation. 
Black holes lose their mass by emitting the Hawking radiation 
and become as small as the Planck length in the last stages. 
The Hawking temperature of the black hole becomes also comparable to the Planck scale, 
and quantum effects of gravity will be important. 
String theory is a promising candidate of quantum theory of gravity 
and would provide a better description of the final state of black hole evaporation. 

It has been proposed that small black holes would transit into strings 
when the size of the black hole becomes comparable 
to the string scale \cite{Bowick:1985af, Susskind:1993ws, Horowitz:1997jc, Chen:2021dsw, Susskind:2021nqs}. 
Unfortunately, it is very difficult to describe the dynamical process 
around the end of black hole evaporation by using string theory. 
Instead, some sort of the phase diagram of static states of black holes 
and strings have been studied. 
Susskind proposed the correspondence between black holes and fundamental string states 
by considering adiabatic change of the string coupling \cite{Susskind:1993ws} 
(See also \cite{Susskind:2021nqs} for a brief review). 
As the string coupling increases, a highly excited string will shrink its size 
by the self-gravitation and will eventually be smaller than its Schwarzschild radius. 
Then, the highly excited string must turn into a black hole. 

Horowitz and Polchinski developed Susskind's idea further and formulated a model of 
the self-gravitating strings \cite{Horowitz:1997jc}.
\footnote{%
For related works, see, for example, \cite{Horowitz:1996nw, Damour:1999aw, Khuri:1999ez, Barbon:2004dd, Giveon:2006pr, Mertens:2014dia, Kawamoto:2015zha, Brustein:2021ifl, Chen:2021emg, Balthazar:2022szl, Brustein:2022uft}.
}
They considered an effective field theory of winding strings 
wrapping on the Euclidean time circle. 
String theory has a temperature called the Hagedorn temperature, 
beyond which the thermal ensemble would not be well-defined \cite{Atick:1988si}. 
Horowitz and Polchinski studied winding strings near the Hagedorn temperature 
and found that there is a solution of bound states of strings by the self-gravitation. 

In this paper, we study the transition between black holes and strings in more details. 
In order to see what happens around the transition, 
non-linear effects of gravity should be taken into account. 
Around the Hagedorn temperature, strings behave as almost free strings. 
As the temperature decreases, interactions between strings become more important, 
and strings form a bound state by self-gravitation. 
Gravitation between strings is weak as long as the temperature 
is sufficiently close to the Hagedorn temperature. 
Horowitz and Polchinski studied the bound state in this regime by using the linearized gravity. 
The transition to a black hole occurs at a still lower temperature. 
As the temperature decreases, the size of the bound state approaches the Schwarzschild radius, 
and gravitation becomes stronger. 
Naively, strings are expected to fall inside the Schwarzschild radius 
at some critical temperature, and turn into a black hole. 
Around the critical temperature, gravitation is very strong, as the event horizon is almost formed. 
In order to describe the bound state in this regime, 
non-linear effects of gravity should be taken into account. 
Thus, the transition cannot be studied by using the approximation of the linearized gravity. 

In this paper, we approximate the bound state of strings by a perfect fluid of winding strings, 
and take the non-linear gravity into calculations by solving 
the Einstein equation for the winding string fluid. 
Since fluid is one of the most typical states of local thermal equilibrium, 
it is natural to expect that winding strings near the Hagedorn temperature behave as a fluid, 
and dissipation would be negligible in static configurations of the bound state.  
Near the Hagedorn temperature, our fluid solution reproduces 
results of the Horowitz-Polchinski model. 
The local temperature inside the bound state is still very high because of the blue-shift
even at lower temperatures from the viewpoint of a fiducial observer. 
The solution can be continued to lower temperatures and 
approaches the Schwarzschild black hole. 
Thus, our fluid model provides a description of the transition 
between black holes and self-gravitating strings. 
The solution behaves as a bound state of strings at high temperatures and 
becomes a black hole approximately at low temperatures. 
Moreover, it is expressed as a smooth function of the temperature 
implying that two phases are continuously connected to each other, 
contrary to the naive expectation that 
strings would collapse into a black hole at the critical temperature. 

The paper is organized as follows. 
In Sec.~\ref{sec:HP}, we briefly review the Horowitz-Polchinski model. 
In Sec.~\ref{sec:Fluid}, we show that winding strings approximately behave as a perfect fluid. 
In Sec.~\ref{sec:Geometry}, we solve the Einstein equation for the fluid of winding strings 
to obtain the geometry of the star of winding strings. 
In Sec.~\ref{sec:Limits}, we consider two special cases of 
the high temperature limit and low temperature limit. 
In Sec.~\ref{sec:Buchdahl}, we discuss why our solution can violate 
the Buchdahl inequality in the low temperature limit. 
In Sec.~\ref{sec:Higher}, we show the results for higher dimensions. 
Sec.~\ref{sec:Conclusion} is devoted for the conclusion and discussions.


\section{Horowitz-Polchinski model}\label{sec:HP}

In this section, we briefly review the Horowitz-Polchinski model \cite{Horowitz:1997jc}, 
which describes self-gravitating strings near the Hagedorn temperature \cite{Atick:1988si}. 
Strings at a finite temperature can be studied by using 
the Euclidean spacetime with the time direction compactified to a circle. 
Strings propagating in the Euclidean time direction 
can be viewed as winding strings wrapping the Euclidean time circle. 
Before taking the GSO projection, the lowest mode of the closed string is tachyonic, 
\begin{equation}
 m^2_0 = - \frac{\beta_H^2}{(2\pi \alpha')^2} \ , 
\end{equation}
where $\beta_H$ is the inverse Hagedorn temperature, which is given by 
$\beta_H = 4 \pi \alpha^{\prime\,1/2}$ for bosonic strings or 
$\beta_H = 2^{3/2} \pi \alpha^{\prime\,1/2}$ for Type II strings. 
Winding strings wrapping the Euclidean time circle with an inverse temperature $\beta$ 
have additional contribution to the mass due to the string tension, 
\begin{equation}
 m^2 = \frac{\beta^2 - \beta_H^2}{(2\pi \alpha')^2} \ . 
 \label{mass}
\end{equation}
The GSO projection does not exclude the tachyonic mode 
due to the anti-periodic boundary condition in the Euclidean time circle for fermionic states \cite{Atick:1988si}. 
Thus, the tachyonic mode appears in the winding strings when 
the temperature exceeds the Hagedorn temperature, $\beta < \beta_H$. 
The tachyonic instability in this regime implies that 
the thermal ensemble is no longer well-defined.  
Horowitz and Polchinski showed that the winding condensate 
occurs even below the Hagedorn temperature due to the redshift by the self-gravitation. 

Horowitz and Polchinski considered the effective field theory of winding strings near the Hagedorn temperature. 
After the Kaluza-Klain reduction of the Euclidean time circle, 
the action of the effective field theory in $(D-1)$-dimensional space is given by 
\begin{equation}
 \mathcal I 
 = 
 \frac{\beta}{16\pi G_N} \int d^{D-1}x \sqrt{g_{D-1}}\, e^{-2\Phi_{D-1}} 
 \left[
 - R_{D-1}  - 4 \left(\partial\Phi_{D-1}\right)^2 + (\partial\varphi)^2 
 + \left|\partial\chi\right|^2 + m^2_\text{eff}(\varphi) \left|\chi\right|^2 
 \right] \ , 
 \label{ActionE}
\end{equation}
where $\chi$ is the complex field of winding strings. 
Here, we consider only configurations without the Kaluza-Klein charge 
and ignore the Kaluza-Klein gauge field. 
The redshift factor, or equivalently, the Euclidean time component of the metric is 
\begin{equation}
 g_{\tau \tau} = e^{2\varphi} \ , 
\end{equation}
and the $(D-1)$-dimensional dilaton field $\Phi_{D-1}$ is related to 
the original $D$-dimensional dilaton field $\Phi$ 
as 
\begin{equation}
 \Phi_{D-1} = \Phi - \frac{1}{2} \varphi \ . 
\label{dilaton}
\end{equation}
The local radius of the Euclidean time circle becomes smaller 
due to the blue-shift, and the effective mass of winding strings is given by 
\begin{equation}
 m^2_\text{eff}(\varphi) 
 = 
 \frac{e^{2\varphi}\beta^2 - \beta_H^2}{(2\pi \alpha')^2} \ . 
 \label{meff}
\end{equation}

Now, we consider the equation of motion of \eqref{ActionE}. 
As the temperature is very close to the Hagedorn temperature, 
we expand the equation of motion to the leading order of  
\begin{equation}
 \frac{\beta^2 - \beta_H^2}{2\pi \alpha'} \equiv \epsilon \ . 
\end{equation}
We also make the ansatz that $\varphi$ is of the same order; 
\begin{equation}
 \frac{\beta^2}{2\pi \alpha'} \varphi \simeq \frac{\beta_H^2}{2\pi \alpha'} \varphi \sim \mathcal O(\epsilon) \ , 
 \label{weak}
\end{equation}
so that the effective mass \eqref{meff} 
is of the same order to the mass in the flat space \eqref{mass}. 
Then, the equations of motion at the leading order in small $\epsilon$ become 
\begin{align}
 0 &= 
 \nabla^2 \chi 
 - m^2_\text{eff}(\varphi) \chi 
 \ , \label{eq-chi}
\\
 0 &= 
 \nabla^2 \varphi - \frac{\beta_H^2}{(2\pi \alpha')^2} \left|\chi\right|^2
 \ ,  \label{eq-phi}
\end{align}
where the effective mass $m^2_\text{eff}(\varphi)$ at the leading order is 
\begin{equation}
 m^2_\text{eff}(\varphi) \simeq  
 \frac{\beta^2 - \beta_H^2}{(2\pi \alpha')^2} 
 + \frac{2 \varphi\beta_H^2}{(2\pi \alpha')^2} \ . 
\end{equation}
The metric is approximately flat and the dilaton is trivial to this order; 
\begin{align}
 \left(R_{D-1}\right)_{ij} &= \mathcal O(\epsilon^3) \ , 
 &
 \partial^2 \Phi_{D-1} &= \mathcal O(\epsilon^3) \ . 
\end{align}
The equations of motion \eqref{eq-chi} and \eqref{eq-phi} 
have only one parameter $\epsilon$ and one unit of length $\alpha^{\prime\,1/2}$. 

The equations of motion \eqref{eq-chi} and \eqref{eq-phi} can be rewritten 
into a single non-linear equation of the winding string field $\chi$. 
By solving \eqref{eq-phi}, $\varphi$ is given in the form of integral as 
\begin{equation}
 \varphi(x) 
 = 
 - \frac{\beta_H^2}{(2\pi \alpha')^2 (D-3) \Omega_{D-2}} 
 \int d^{D-1} y \frac{\left|\chi(y)\right|^2}{\left|\vec x - \vec y\right|^{D-3}} 
 \ , \label{eq-HP}
\end{equation}
where $\Omega_{D-2}$ is the area of the unit $(D-2)$-sphere. 
Substituting \eqref{eq-HP} into \eqref{eq-chi}, we obtain 
\begin{align}
 - \nabla^2 \chi(x) 
 - \frac{2 \beta_H^4}{(2\pi \alpha')^4 (D-3) \Omega_{D-2}} 
 \int d^{D-1} y \frac{\left|\chi(y)\right|^2}{\left|\vec x - \vec y\right|^{D-3}} \ \chi(x) 
 &= 
 - \frac{\beta^2 - \beta_H^2}{(2\pi \alpha')^2} \chi(x) \ .  
\end{align}
This integro-differential equation can be viewed as 
a non-linear Schr\"odinger equation with an attractive Coulomb self-interaction. 
In order to see this equation as a non-linear eigenvalue problem, 
it is convenient to introduce rescaled variables as 
\begin{align}
 \hat x 
 &= 
 \sqrt{\frac{\beta^2 - \beta_H^2}{(2\pi \alpha')^2\zeta}}\, x \ , 
 & 
 \hat \chi 
 &= 
 \frac{\sqrt{2}\,\beta_H^2\zeta}{\left(\beta^2 - \beta_H^2\right)\sqrt{(D-3)\Omega_{D-2}}}\, \chi \ , 
 \label{scaling}
\end{align}
where $\zeta$ is chosen so that $\hat\chi$ is normalized as 
\begin{equation}
 \int d^{D-1} \hat x \, \left|\hat\chi\right|^2 = 1 \ . 
 \label{norm}
\end{equation}
The value of $\zeta$ is not determined by this normalization condition 
but obtained by solving the eigenvalue problem of the non-linear Schr\"odinger equation, 
\begin{align}
 - \hat{\nabla}^2 \hat\chi(\hat x) 
 - \int d^{D-1} \hat y \frac{\left|\hat\chi(\hat y)\right|^2}{\left|\hat x - \hat y\right|^{D-3}} \ 
 \hat\chi(\hat x) 
 &= 
 - \zeta \hat\chi(\hat x) \ . 
 \label{eq-chi-HP}
\end{align}
Since the interaction is attractive, this Schr\"odinger equation has 
normalizable solutions of bound states with discrete eigenvalues of $\zeta$. 
Bound states of self-gravitating strings would be described by the ``ground state'' 
with maximal value of $\zeta$.%
\footnote{%
The eigenvalue $\zeta$ is not the energy of the winding string field $\chi$, 
and hence, the ground state may not have minimum energy in solutions. 
However, the ground state would be the most uniform configuration as the wave function has no node ---  
$|\chi|^2$ is positive everywhere except for the spatial infinity. 
Thus, the ground state is expected to give a best approximation of the bound state. 
}
Since the solution is given in terms of the rescaled coordinate $\hat x$, 
it must scale as 
\begin{equation}
 \ell \sim \sqrt{ \frac{2\pi \alpha'}{\epsilon}} = \frac{2\pi \alpha'}{\sqrt{\beta^2 - \beta_H^2}} \ .  
 \label{size}
\end{equation}

Since the winding strings describe strings at finite temperatures, 
the entropy $S$ can also be estimated by using the first law of thermodynamics; 
\begin{equation}
 S 
 = \left(\beta \partial_\beta - 1\right) \mathcal I \ . 
 \label{Entropy0}
\end{equation}
The entropy can be calculated by taking the derivative only of 
the explicit $\beta$-dependence in \eqref{ActionE} as%
\footnote{%
For the other implicit $\beta$-dependence in the fields, 
the variation 
vanishes since the solution satisfies the equation of motion. 
}
\begin{equation}
 S 
 = 
 \int d^{D-1}x \sqrt{g_{D-1}}\, e^{-2\Phi} 
 \frac{e^{3\varphi}\beta^3}{8\pi G_N (2\pi \alpha')^2} \left|\chi\right|^2 \ . 
 \label{Entropy}
\end{equation}
By using \eqref{scaling} and \eqref{norm}, 
the entropy of the solution is expressed as 
\begin{equation}
 S 
 \simeq 
 \frac{(D-3) (2\pi \alpha')^{D-3} \zeta^{\frac{D-5}{2}} \Omega_{D-2}}
 {16\pi G_N \beta_H \left(\beta^2-\beta_H^2\right)^{\frac{D-5}{2}}} \ . 
\end{equation}
The mass of the bound state is estimated as 
\begin{equation}
 M \simeq \beta_H^{-1} S 
 = 
 \frac{(D-3) (2\pi \alpha')^{D-3} \zeta^{\frac{D-5}{2}} \Omega_{D-2}}
 {16\pi G_N \beta_H^2 \left(\beta^2-\beta_H^2\right)^{\frac{D-5}{2}}} \ . 
 \label{MassHP}
\end{equation}

Although we obtained a single Schr\"odinger equation for the winding string field $\chi$, 
it might be technically easier to solve the original set of differential equations 
\eqref{eq-chi} and \eqref{eq-phi} than the integro-differential equation \eqref{eq-chi-HP}, in practice. 
The original equations have no free parameter that can be treated as the eigenvalue. 
As we have seen, the eigenvalue $\zeta$ comes from the normalization condition of $\hat\chi$, 
implying that the normalization of $\chi$ cannot be fixed by hand 
but is automatically determined when we solve the differential equations \eqref{eq-chi} and \eqref{eq-phi}. 
Eqs.~\eqref{eq-chi} and \eqref{eq-phi} are difficult to be solved analytically, 
but can be solved numerically by imposing boundary conditions at $r=0$ and $r\to\infty$ 
so that the solution is normalizable and non-singular. 

It should be noted that the solution above is valid only near the Hagedorn temperature. 
Since the Horowitz-Polchinski model relies on the ansatz \eqref{weak} 
and the equations of motion \eqref{eq-chi} and \eqref{eq-phi} 
are approximated by the linearized gravity. 
In order for the validity of the approximation, gravitation must be sufficiently weak. 
The size of the bound state becomes smaller as the temperature is lowered, 
and becomes comparable to the Schwarzschild radius at some temperature. 
Then, gravity around the bound state is very strong, and the Horowitz-Polchinski model 
cannot give a good description of the bound state. 

The Schwarzschild radius of the bound state is given in terms of the mass \eqref{MassHP} as 
\begin{equation}
 r_h = \left(\frac{16\pi G_N M}{(D-2)\Omega_{D-2}}\right)^{\frac{1}{D-3}} 
 \sim \frac{2\pi \alpha'}{\beta_H^{ \frac{2}{D-3}}\left(\beta^2-\beta_H^2\right)^{\frac{D-5}{2(D-3)}}} \ , 
\end{equation}
up to some numerical factor. 
Thus, the size of the bound state becomes comparable to the Schwarzschild radius 
at a temperature lower than but still comparable with the Hagedorn temperature, 
\begin{equation}
 \frac{\beta^2 - \beta_H^2}{\beta_H^2} = \mathcal O(1) \ . 
\end{equation}
Thus, the solution can be invalid even at a temperature comparable with the Hagedorn temperature. 
For the validity of the solution, the temperature must be very close to the Hagedorn temperature as%
\footnote{%
Thermal states are approximated by a solution of the equation of motion if 
if quantum fluctuations are suppressed. The action is roughly estimated as 
\begin{equation}
 \mathcal I \sim \frac{(D-3) (2\pi \alpha')^{D-3} \zeta^{\frac{D-5}{2}} \Omega_{D-2}}
 {16\pi G_N \beta_H^3 \left(\beta^2-\beta_H^2\right)^{\frac{D-7}{2}}} \ , 
\end{equation}
and it should be sufficiently large for the suppression of fluctuations. 
Thus, a solution gives a good description for 
\begin{equation}
 \left(\frac{\beta^2-\beta_H^2}{\beta^2}\right)^{\frac{7-D}{2}} 
 \gtrsim 
 {\alpha'}^{ \frac{2-D}{2}} G_N
 \ . 
\end{equation}
This condition gives an upper bound of the temperature for $D<7$. 
The temperature should be close to the Hagedorn temperature but should not be too close to it. 
}
\begin{equation}
 \beta^2 - \beta_H^2 \ll \beta_H^2 \ . 
\end{equation}

In this paper, we study the transition between black holes and strings. 
Gravitation becomes very strong near the critical point and 
cannot be described by the linearized gravity. 
We will make a simplification of the problem 
to take non-linear effects of gravity into account. 
In the next section, we first introduce an approximation of winding strings by a perfect fluid. 
Then, we derive the analytic solution of the fluid model in the subsequent sections.

%
%
\section{Fluid approximation of winding strings}\label{sec:Fluid}

In this section, 
we introduce our fluid approximation of winding strings. 
We propose that the stress-energy tensor of winding strings 
can be approximated by 
\begin{align}
 T_{tt} &= - \rho\, g_{tt} \ , 
 & 
 T_{ij} &= P\, g_{ij} \ , 
 &
 T_{ti} &= 0 \ , 
 \label{fluidT}
\end{align}
with the energy density $\rho$ and pressure $P$ given by 
\begin{align}
 \rho 
 &= 
 e^{-2\Phi}\frac{3 e^{2\varphi}\beta^2 - \beta_H^2}{16\pi G_N (2\pi \alpha')^2} \left|\chi\right|^2 \ , 
 &
 P
 &= 
 e^{-2\Phi}\frac{\beta^2_H - e^{2\varphi}\beta^2}{16\pi G_N (2\pi \alpha')^2} \left|\chi\right|^2 \ , 
 \label{fluid}
\end{align}
at least in a static bound state. 
Here, the time direction labeled by $t$ is the Lorentzian time 
associated to the time independence of the static configuration. 
Indices $i$ and $j$ indicates the spacial directions on time slices with $t = \mathrm{const}$. 

We derive the stress-energy tensor \eqref{fluidT}--\eqref{fluid} 
in two different ways. 
In Sec.~\ref{ssec:fluid}, we show that contributions from momenta are sufficiently small 
compared with those from the mass of winding strings, 
and then, obtain \eqref{fluidT}--\eqref{fluid} by ignoring the kinetic terms. 
It is reasonable that momenta are negligible in the bound state of winding strings 
since the bound state is described by the ground state, in which strings have only zero point fluctuations. 

In Sec.~\ref{ssec:energy}, we assume that winding strings 
approximately behave as a perfect fluid, and then, 
derive \eqref{fluidT}--\eqref{fluid} from thermodynamic relations. 
In the Horowitz-Polchinski model, thermal states 
near the Hagedorn temperature are studied, and the solution 
involves the condensate of winding strings. 
The bound state of winding strings can be viewed as 
a bound state of many strings which are created by thermal energy. 
Thus, it is expected that winding strings in the bound state can be approximated by a fluid. 
In Sec.~\ref{ssec:energy}, we show that the energy density and pressure 
are always given by \eqref{fluid} if winding strings behave as a perfect fluid. 

It should be noted that the argument in Sec.~\ref{ssec:fluid} 
shows that the stress-energy tensor takes the form of \eqref{fluidT},  
but does not necessarily imply that winding strings behave as a perfect fluid. 
There might be corrections which cannot be seen in static configurations. 
For example, winding strings may not behave as a perfect fluid 
but have non-zero viscosity or higher derivative corrections. 
Also, the typical scale of the fluid may be longer than the size of the bound state. 
For our purpose of studying the transition between black holes and strings, 
it is not very important whether winding strings really behave as a fluid, 
but the expression \eqref{fluidT}--\eqref{fluid} is important. 


\subsection{Approximation of the stress-energy tensor}\label{ssec:fluid}

Here, we consider the stress-energy tensor of the Horowitz-Polchinski model 
and show that it can be approximated by \eqref{fluidT}--\eqref{fluid}. 
The stress-energy tensor is given by the variation of 
the matter part of the action with respect to the metric, 
\begin{equation}
 T^{\mu\nu} = \frac{2}{\sqrt{-g}} \frac{\delta \mathcal I_\text{matter}}{\delta g_{\mu\nu}} \ , 
\end{equation}
where $\mathcal I_\text{matter}$ is the matter part of the action. 
In order to calculate the stress-energy tensor, 
we consider the $D$-dimensional field theory which gives \eqref{ActionE} 
after the Wick rotation and the dimensional reduction of the Euclidean time circle. 
The action is given by 
\begin{equation}
 \mathcal I 
 = 
 \frac{1}{16\pi G_N} \int d^D x \sqrt{-g} \, 
 e^{-2\Phi}\left[
 R + 4 \left(\partial\Phi\right)^2 
 - \left|\partial\chi\right|^2 
 + \frac{\beta^2 g_{tt} + \beta^2_H}{(2\pi\alpha')^2}\left|\chi\right|^2
 \right] \ , 
 \label{Action}
\end{equation}
where $\Phi$ is the dilaton in $D$-dimensions, which is related to $\Phi_{D-1}$ by \eqref{dilaton}. 
The stress-energy tensor of winding strings for this model is given by 
\begin{align}
 16\pi G_N \, e^{2\Phi} \, T_{tt} 
 &= 
 g_{tt}\frac{3 \beta^2 g_{tt} + \beta^2_H}{(2\pi\alpha')^2}\left|\chi\right|^2 
 - g_{tt} g^{rr} \left|\partial_r\chi\right|^2 \ , 
 \label{Ttt}
 \\
 16 \pi G_N \, e^{2\Phi} \, T_{rr} 
 &= 
 g_{rr}\frac{\beta^2 g_{tt} + \beta^2_H}{(2\pi\alpha')^2}\left|\chi\right|^2 
 + \left|\partial_r\chi\right|^2 \ , 
 \label{Trr}
 \\
 16 \pi G_N \, e^{2\Phi} \, T_{\theta\theta} 
 &= 
 g_{\theta\theta}\frac{\beta^2 g_{tt} + \beta^2_H}{(2\pi\alpha')^2}\left|\chi\right|^2 
 - g_{\theta\theta} g^{rr}\left|\partial_r\chi\right|^2 \ , 
 \label{Tyy}
\end{align}
and off-diagonal components vanish. 
Here, we considered winding strings with mass \eqref{mass}, which have no Kaluza-Klein momentum 
in the Euclidean time direction, and focused on configurations without angular momentum. 
The index of $\theta$ indicates any of the angular directions.

In the Horowitz-Polchinski model, we consider the linear order approximation 
around the Hagedorn temperature and focus on configurations in which fields are very small. 
In this limit, the effective mass behaves as 
\begin{equation}
 \alpha' m^2_\text{eff}(\varphi) = \mathcal O(\epsilon) \ . 
\end{equation}
Then, eq.~\eqref{eq-phi} implies that 
\begin{equation}
 \alpha'\nabla^2 \sim \mathcal O(\epsilon) \ , 
\end{equation}
for $\chi$. 
Assuming that the Laplacian has a similar scaling behavior for $\varphi$, 
eq.~\eqref{eq-phi} with \eqref{weak} gives 
\begin{equation}
 \alpha'{}^{ \frac{D-2}{4}}\chi = \mathcal O(\epsilon) \ . 
\end{equation}
We further focus on spherically symmetric solutions, 
and the Laplacian is expressed in terms of the radial coordinate in the flat space $r$ as 
\begin{equation}
 \nabla^2 = \partial_r^2 - \frac{D-2}{r} \partial_r \ . 
 \label{lap}
\end{equation}
The solutions have a typical size of $\ell \sim \alpha^{\prime\,1/2} \epsilon^{-1/2}$. 
Near the surface of the solution, the derivative has the typical behavior of 
\begin{equation}
 \partial_r \sim \mathcal O(\epsilon^{1/2}) \ . 
\end{equation}
In the distribution of winding strings except near the surface of the solution, 
or equivalently for $r\ll \ell$, 
the second term of \eqref{lap} will be more important. 
For $r\sim \mathcal O(\epsilon)$, we have 
\begin{equation}
 \nabla^2 \sim - \frac{D-2}{r} \partial_r = \mathcal O(\epsilon) \ . 
\end{equation}
Thus, inside the ``star'' of winding strings, the derivative behaves as 
\begin{equation}
 \partial_r \sim \mathcal O(\epsilon) \ . 
 \label{del-r}
\end{equation}

Now, we show that the stress-energy tensor \eqref{Ttt}--\eqref{Tyy} 
is approximated by \eqref{fluidT}--\eqref{fluid}. 
The mass term of the winding string field behaves as 
\begin{equation}
 m_\text{eff}^2 \left|\chi\right|^2 = \mathcal O(\epsilon^3) \ , 
 \label{meff-order}
\end{equation}
while the kinetic term give only higher order corrections, 
\begin{equation}
 \left|\partial_r\chi\right|^2 = \mathcal O(\epsilon^4) \ . 
 \label{kin-order}
\end{equation}
Then, the stress-energy tensor at the leading order in the small-$\epsilon$ expansion becomes 
\begin{align}
 16\pi G_N \, e^{2\Phi} \, T_{tt} 
 &= 
 - g_{tt}\frac{3 \beta^2 e^{2\varphi} - \beta^2_H}{(2\pi\alpha')^2}\left|\chi\right|^2 
 + \mathcal O(\epsilon^4) \ , 
 \label{Ttt-fluid}
 \\
 16 \pi G_N \, e^{2\Phi} \, T_{ij} 
 &= 
 g_{ij}\frac{\beta^2_H - e^{2\varphi}\beta^2}{(2\pi\alpha')^2}\left|\chi\right|^2 
 + \mathcal O(\epsilon^4) \ . 
 \label{Tij-fluid}
\end{align}
Thus, the stress-energy tensor \eqref{Ttt}--\eqref{Tyy} is approximated by \eqref{fluidT}--\eqref{fluid}. 

So far, we assumed that the temperature is very close to the Hagedorn temperature. 
Discussions above can be generalized to lower temperatures, 
since the local temperature inside the star always exceeds the Hagedorn temperature. 
In order to see the structure of the star of winding strings, 
it is convenient to consider eq.~\eqref{eq-chi} for a given gravitational potential $\varphi$. 
Then, eq.~\eqref{eq-chi} can be interpreted as the Schr\"odinger equation; 
\begin{equation}
 \nabla^2 \chi(r) - V(r) \chi(r) = - E \chi(r) 
\end{equation}
with 
\begin{align}
 V(r) &= - \frac{e^{2\varphi}\beta^2}{(2\pi \alpha')^2} \ , 
 & 
 E &= - \frac{\beta_H^2}{(2\pi \alpha')^2} \ . 
\end{align}
Normalizable eigenstates can be obtained by tuning $\beta$ so that 
the inverse Hagedorn temperature gives the eigenvalue. 
The winding string field $\chi$ is trapped up to quantum penetration 
inside the classical turning point, or equivalently in the region $E>V(r)$. 
The local temperature is blue-shifted and given by the Tolman temperature, 
\begin{equation}
 T_\text{local} = e^{-\varphi} \beta^{-1} \ , 
\end{equation}
which exceeds the Hagedorn temperature for $E>V(r)$. 
Thus, the local temperature is always close to the Hagedorn temperature even for $\beta\gg \beta_H$,%
\footnote{%
To be more precise, the effective mass inside the winding condensate is always small 
compared with the typical scale of the model. 
It can be confirmed straightforwardly from the concrete solution in the subsequent sections 
but should be considered as an ansatz in this sense. 
}
and hence, the stress-energy tensor is approximated by \eqref{fluidT}--\eqref{fluid} 
even at lower temperatures.%
\footnote{%
For $\beta\gg \beta_H$, eq.~\eqref{eq-phi} should be modified by non-linear effects of gravity. 
The winding string field is no longer small as $\chi = \mathcal O(\epsilon^0)$. 
Then, we can see only from eq.~\eqref{eq-chi} that 
\eqref{meff-order} and \eqref{kin-order} become $m_\text{eff}^2 \left|\chi\right|^2 = \mathcal O(\epsilon)$ 
and $\left|\partial_r\chi\right|^2 = \mathcal O(\epsilon^2)$. 
}


The approximation \eqref{Ttt-fluid}--\eqref{Tij-fluid} can also be understood as follows. 
The bound state of winding strings is given by the ground state of 
the equation of motion \eqref{eq-chi} and \eqref{eq-phi}. 
The ground state has only the zero-point fluctuation around the bottom of the potential 
and the kinetic term is much smaller than the potential energy if the potential energy is non-zero. 
Thus, we can ignore the kinetic term as an approximation and obtain \eqref{Ttt-fluid}--\eqref{Tij-fluid}. 

In fact, 
the condition \eqref{del-r} is good only 
for the ground state of the winding string field $\chi$. 
Size of the excited state with the principle quantum number $n$ would be estimated as 
\begin{equation}
 \ell \sim \alpha^{\prime\,1/2} \epsilon^{-1/2} n \ , 
\end{equation}
and hence, the region $r\sim \alpha^{\prime\,1/2} \epsilon^{-1/2}$ would be more important 
than the interior $r\ll \alpha^{\prime\,1/2} \epsilon^{-1/2}$. 
In fact, the momentum would not be small in excited states. 
Highly excited states can be approximated well by the WKB approximation 
and momentum satisfies 
\begin{equation}
 - \nabla^2 \sim p^2 \simeq - m^2_\text{eff}(\varphi) \ , 
\end{equation}
inside the classical turning point. 
 
On the other hand, momentum of the ground state approaches zero in the classical limit, 
as the energy of the classical winding string would simply be given by its mass. 
Small momentum of the ground state comes from the zero point fluctuation or quantum penetration. 
Breakdown of the approximation around the surface of the ``star,'' 
or equivalently the classical turning point, 
would be interpreted as the effect of the quantum penetration, 
which would be absent in the classical limit. 
Thus, 
the approximation of the stress-energy tensor of winding strings \eqref{fluidT}--\eqref{fluid} 
can be understood as some sort of the classical approximation.


\subsection{Derivation from thermodynamic relations}\label{ssec:energy}

In Sec.~\ref{ssec:fluid}, we have seen that the stress-energy tensor of 
winding strings can be approximated by \eqref{fluidT}--\eqref{fluid} for 
the solution of the equations of motion for the ground state. 
Here, we show an alternative derivation of \eqref{fluidT}--\eqref{fluid} 
by using the assumption that winding strings behave as a perfect fluid, 
instead of specific structures of the equation of motion. 
In the bound state of winding strings, the local temperature exceeds 
the Hagedorn temperature due to the blue-shift. 
The winding strings are condensing inside the bound state, 
and many strings are excited by thermal energy in the Lorentzian picture. 
Thus, it is natural to expect that winding strings in the bound state behave as a fluid. 
Although the fluid may not be a perfect fluid 
but possibly have some corrections such as the viscosity, 
we consider a perfect fluid as a simplest model assuming that 
corrections are sufficiently small and negligible at least in static configurations. 


Here, we show two different but essentially equivalent procedures 
by using the assumption that winding strings behave as a perfect fluid. 
First, we calculate the entropy and total energy (i.e.\ the ADM mass) 
from the free energy (i.e.\ the action) by using thermodynamic relations. 
Directly from the assumption, the stress-energy tensor takes the form of \eqref{fluidT}. 
Then, the energy density and pressure can be read off from 
the entropy and the ADM mass, and turn out to be given by \eqref{fluid}. 

In another method, we consider the stress-energy tensor of winding strings. 
By assuming that winding strings behave as a perfect fluid, 
kinetic terms should be isotropic, but still can be non-zero. 
We show that the kinetic term must vanish to satisfy the local thermodynamic relation, 
and then, the stress-energy tensor becomes \eqref{fluidT}--\eqref{fluid}. 
In both of two procedures, 
we do not resort to any specific structure of the solution, 
but just assume that the fluid is in local thermal equilibrium and static. 
Hence, winding string field $\chi$ is not necessarily in the ground state. 


First, we calculate the energy density and pressure from the entropy and the ADM mass. 
The entropy of winding strings can be calculated from the action 
by using the first law of thermodynamics \eqref{Entropy0} and is obtained as \eqref{Entropy}. 
We assume that the integrand of the total entropy \eqref{Entropy} is the local entropy density $s$, 
which can be read off as 
\begin{equation}
 s 
 = 
 e^{-2\Phi} \frac{e^{3\varphi}\beta^3}{8\pi G_N (2\pi \alpha')^2} \left|\chi\right|^2 \ . 
 \label{s}
\end{equation}
%
%
Assuming that winding strings behave as a perfect fluid, 
the entropy density satisfies the local thermodynamic relation 
with the energy density and pressure \eqref{fluid} as 
\begin{equation}
 s = e^{\varphi}\beta \left(\rho + P\right) \ , 
 \label{thermo}
\end{equation}
where $e^{\varphi} \beta$ is the local inverse temperature. 

By using the thermodynamic relation with the total action, 
we obtain the energy of the total system, which is nothing but the ADM mass;%
\footnote{%
The additional factor of $2(D-3)/(D-2)$ appears because 
the ADM mass in higher dimensions is defined so that 
the coefficient of $\rho$ in \eqref{ADM0} becomes $1$ and 
is different from the Komar integral by this factor. 
} 
\begin{equation}
 \frac{2(D-3)}{D-2} M = - \partial_\beta \mathcal I \ . \label{ADM-thermo}
\end{equation}
As the ADM mass contains contributions from the gravity part, 
it is not given by the integration of the energy density of winding strings.%
\footnote{%
If we use the matter part of the action instead of the total action, 
the thermodynamic relation gives the energy of matters, 
which is the integration of the energy density. 
}
By using the equations of motion, 
the ADM mass can be expressed in terms of the stress-energy tensor. 
When the stress-energy tensor takes the form of the perfect fluid in the Einstein frame, 
the ADM mass is given in terms of the energy density and pressure as 
\begin{equation}
 M = 
 \int d^{D-1}x \sqrt{-g} 
 \left(\rho + \frac{D-1}{D-3} P\right) \ . 
 \label{ADM0}
\end{equation}
By using the formula \eqref{ADM-thermo} with equations of motion, 
we obtain the following expression of the ADM mass; 
\begin{align}
 M 
 &= 
  \int d^{D-1}x \sqrt{-g}\, e^{-2\Phi} 
 \frac{\beta_H^2 + (D-4) e^{2\varphi}\beta^2}{8(D-3)\pi G_N (2\pi \alpha')^2} 
  \left|\chi\right|^2 \ . 
 \label{ADM}
\end{align}
By comparing two expressions \eqref{ADM0} and \eqref{ADM}, 
we obtain 
\begin{equation}
 \rho + \frac{D-1}{D-3} P 
 = 
 e^{-2\Phi} 
 \frac{\beta_H^2 + (D-4) e^{2\varphi}\beta^2}{8(D-3)\pi G_N (2\pi \alpha')^2} 
  \left|\chi\right|^2 \ . 
\end{equation}
Together with the condition \eqref{thermo} with \eqref{s}, 
the energy density and pressure can be calculated, 
and we obtain \eqref{fluid}. 



Alternatively, we can show that kinetic terms in the stress-energy tensor 
must vanish if winding strings behave as a perfect fluid,%
\footnote{%
Of course, kinetic terms vanish only for the winding string field $\chi$, 
and do not vanish if the fluid consists of ordinary matter fields. 
} 
by using thermodynamic relations, to obtain \eqref{fluid}. 
The assumption that matters behave as a perfect fluid 
means that the pressure is isotropic, or equivalently that 
the spatial components of the stress-energy tensor is proportional to the unit matrix. 
This assumption indicates that contributions from the kinetic terms must satisfy 
\begin{equation}
 \left| \partial_i\chi\right|^2 = \left| \partial_j\chi\right|^2 = \frac{1}{D-1}\left| \nabla\chi\right|^2 \ , 
\end{equation} 
for any spatial directions $i$ and $j$ in the local lorentz frame. 
Then, the stress-energy tensor of the winding strings becomes  
\begin{align}
 16\pi G_N \, e^{2\Phi} \, T_{tt} 
 &= 
 g_{tt}\frac{3 \beta^2 g_{tt} + \beta^2_H}{(2\pi\alpha')^2}\left|\chi\right|^2 
 - g_{tt} \left|\nabla\chi\right|^2 \ , 
 \label{Ttt-F}
 \\
 16 \pi G_N \, e^{2\Phi} \, T_{ii} 
 &= 
 g_{ii}\frac{\beta^2 g_{tt} + \beta^2_H}{(2\pi\alpha')^2}\left|\chi\right|^2 
 + 2 \left|\partial_i\chi\right|^2 - g_{ii} \left|\nabla\chi\right|^2 \ , 
\notag\\
 &= 
 g_{ii}\frac{\beta^2 g_{tt} + \beta^2_H}{(2\pi\alpha')^2}\left|\chi\right|^2 
 - \frac{D-3}{D-1} g_{ii} \left|\nabla\chi\right|^2 \ . 
 \label{Tii-F}
\end{align}
Then, the energy density and pressure of the fluid are read off as 
\begin{align}
 \rho 
 &= 
 e^{-2\Phi}\frac{3 e^{2\varphi}\beta^2 - \beta_H^2}{16\pi G_N (2\pi \alpha')^2} \left|\chi\right|^2 
 + \frac{1}{16\pi G_N} e^{-2\Phi} \left|\nabla\chi\right|^2 \ , 
\\
 P
 &= 
 e^{-2\Phi}\frac{\beta^2_H - e^{2\varphi}\beta^2}{16\pi G_N (2\pi \alpha')^2} \left|\chi\right|^2 
 - \frac{D-3}{16(D-1)\pi G_N} e^{-2\Phi} \left|\nabla\chi\right|^2 \ . 
 \label{fluid-thermo}
\end{align}
By using the local thermodynamic relation \eqref{thermo}, 
we obtain the entropy density, 
\begin{equation}
 s 
 = 
 e^{-2\Phi} \frac{e^{3\varphi}\beta^3}{8\pi G_N (2\pi \alpha')^2} \left|\chi\right|^2 
 + \frac{e^{\varphi}\beta}{8(D-1)\pi G_N} e^{-2\Phi} \left|\nabla\chi\right|^2 \ . 
 \label{s-kin}
\end{equation}
The entropy can also be calculated 
from the thermodynamic relation \eqref{Entropy0}, and then, 
the entropy density is given by \eqref{s}. 
By comparing \eqref{s} and \eqref{s-kin}, we find that 
contributions from the kinetic term vanish; 
\begin{equation}
 \left|\nabla\chi\right|^2 = 0 \ , 
 \label{kin0}
\end{equation}
since the kinetic term in \eqref{s-kin} is non-negative. 
Substituting \eqref{kin0} into \eqref{fluid-thermo}, we obtain \eqref{fluid}. 


Here, we have shown that kinetic terms should vanish 
when winding strings behave as a perfect fluid. 
This is very different from fluids which consist of ordinary matter fields. 
Matters in ordinary fluids usually have non-zero momenta and hence their kinetic terms are non-zero. 
It should be noted that momenta of winding strings cannot be interpreted 
as those of strings in Lorentzian spacetimes before the Wick rotation, 
because winding strings are wrapping on the Euclidean time circle, 
while strings in Lorentzian spacetime do not extend but propagate in the time direction. 

In Sec.~\ref{ssec:fluid}, we have seen that the kinetic term is negligible 
for winding string field $\chi$ in the ground state, 
but this approximation is invalid for $\chi$ in some excited states. 
Thus, for the derivation in Sec.~\ref{ssec:fluid}, 
the ground state should be dominated in the path integral of $\chi$. 
In contrast, the arguments above by using the thermodynamic relation 
is valid as long as winding strings are in local thermal equilibrium and the pressure is isotropic. 
Since we obtained the same result to Sec.~\ref{ssec:fluid}, 
winding strings can be approximated by the solution of the ground state 
when they behave as a fluid. 

At the same time, winding string field $\chi$ in the ground state 
definitely have small but non-zero kinetic terms, 
implying that winding strings cannot be a perfect fluid exactly 
but have some corrections. 
Moreover, the winding string fluid possibly has corrections which cannot be seen in static configurations. 
These corrections can be understood as higher derivative corrections to \eqref{fluidT}--\eqref{fluid}. 
Since the higher derivative corrections for fluids usually give dissipation, 
winding strings may be identified with a dissipative fluid if these corrections are taken into account.%
\footnote{%
Although, most of higher derivative corrections vanish in static configurations, 
some terms may survive in the form of derivatives of metric components. 
}

Apart from the relation to dissipative fluids, 
kinetic terms can be taken into calculations as higher derivative corrections. 
Here, we do not pursue this direction but just ignore these terms. 
In the next section, we study the Einstein equation for 
the stress-energy tensor \eqref{fluidT}--\eqref{fluid}.


\section{Geometry of the winding string fluid}\label{sec:Geometry}

In the previous section, we have seen that 
the winding string field approximately behaves as a perfect fluid. 
Here, we study the static solution of the Einstein equation for this fluid. 
In this and subsequent sections, we ignore the coupling with dilaton and take $\Phi=0$ for simplicity.%
\footnote{%
It should be noted that the dilaton has non-trivial configurations 
due to the coupling with winding strings in the original model. 
Thus, ignoring the dilaton field should be considered as a modification of the model. 
Here, we just assume that the qualitative feature of the solution would not be changed. 
} 
In this section, we also focus on the case of $D=4$. 


\subsection{Solution of the Einstein equation}\label{ssec:solution}

Here, we first consider the interior geometry of the star of winding strings. 
We solve the Einstein equation, 
\begin{equation}
 R_{\mu\nu} - \frac{1}{2}g_{\mu\nu} R = 8\pi G_N T_{\mu\nu} \ , \label{Ein}
\end{equation}
where the stress-energy tensor of the static perfect fluid is given by 
\begin{align}
 T^t{}_t &= -\rho \ , 
 &
 T^r{}_r &= T^\theta{}_\theta = T^\phi{}_\phi = P \ . 
 \label{EMtensor}
\end{align}
As we have seen in Sec.~\ref{ssec:fluid}, 
the energy density and pressure of the winding string fluid are given by \eqref{fluid}. 
The most general metric of spherically symmetric and static spacetimes 
can be expressed up to the coordinate transformation as 
\begin{equation}
 ds^2 = - f(r) dt^2 + \frac{dr^2}{f(r)\,h(r)} + r^2 d \Omega^2 \ , 
 \label{metric}
\end{equation}
where $f(r) = e^{2\varphi(r)}$ and 
\begin{equation}
 d \Omega^2 = d \theta^2 + \sin^2\theta d\phi^2 
\end{equation}
is the metric of unit 2-sphere. 

For the consistency with the Einstein equation, 
the stress-energy tensor must satisfy the conservation law, 
\begin{equation}
 \nabla_\mu T^{\mu\nu} = 0 \ . 
\end{equation}
In this paper, we focus on the spherically symmetric and static configurations. 
Then, the conservation law for the perfect fluid \eqref{EMtensor} on the geometry \eqref{metric} 
gives the following constraint on the energy density and pressure; 
\begin{equation}
 \left(\rho(r) + P(r)\right) f'(r) + 2 P'(r) f(r) = 0 \ . 
 \label{cons}
\end{equation}
By using the energy density and pressure of the winding string fluid \eqref{fluid}, 
the constraint \eqref{cons} becomes 
\begin{equation}
 \frac{d}{dr} \left|\chi\right|^2 = 0 \ , 
\end{equation}
and hence, we define 
\begin{equation}
 \left|\chi\right|^2 \equiv 8 \pi^2 \alpha^{\prime\,2} D_0 = \text{const.} \label{D0-def}
\end{equation}
This condition is consistent with the approximation $\partial_r \sim 0$. 

For static and spherically symmetric geometries, 
only two components of the Einstein equation give independent differential equations. 
From $(t,t)$- and $(r,r)$-components, we obtain 
\begin{align}
 0 
 &= 
 \left[r f(r) h(r)\right]' + r^2 \left(3 f(r) \beta^2 - \beta_H^2\right) D_0 - 1 \ , 
 \\
 0 
 &= 
 h(r) \left[r f(r)\right]' + r^2 \left(f(r) \beta^2 - \beta_H^2\right) D_0 - 1 \ . 
\end{align}
It is straightforward to solve the equations above. 
The solution is obtained as%
\footnote{%
This solution is first considered in \cite{Wyman:1949zz} as early as 1949, 
as a simplest example of spherically symmetric and static solutions of the perfect fluid, 
of course in a different context from string theory. 
} 
\begin{align}
 f(r) 
 &= 
 \frac{\beta_H^2}{\beta^2}\left\{\frac{1}{h_0} + 1  
 - \frac{\sqrt{r_m^2 - r^2}}{r} 
 \left[\sin^{-1}\left(\frac{r}{r_m}\right) + a\right]\right\} \ , 
 \label{f}
 \\
 h(r) 
 &= 
 \frac{\beta^2}{\beta_H^2} h_0 \left(1-\frac{r^2}{r_m^2}\right) \ , 
 \label{h}
\end{align}
where $a$ and $h_0$ are the integration constants. 
The redshift factor $f(r)$ can be real either for $r\leq r_m$ or for $r\geq r_m$, 
depending on the value of the integration constants, 
where 
\begin{equation}
 r_m^2 = \frac{h_0}{\beta_H^2 D_0} \ . 
 \label{rm}
\end{equation}
The solution has the singularity at $r=0$ for $a\neq 0$, 
and hence, we take 
\begin{equation}
 a = 0 \ . \label{a=0}
\end{equation}
Then, $f(r)$ is real only in $r\leq r_m$. 
The solution also satisfies the condition $g_{rr} = 1$ at $r=0$, 
which is necessary to avoid the conical singularity. 
The redshift factor $f(r)$ monotonically increases from 
\begin{equation}
 f(0) =  \frac{\beta_H^2}{\beta^2} \frac{1}{h_0} \ , 
\end{equation}
to 
\begin{equation}
 f(r_m) =  \frac{\beta_H^2}{\beta^2}\left(\frac{1}{h_0} + 1 \right) \ ,  
\end{equation}
as $r$ increases. 
The pressure \eqref{fluid} becomes positive if 
\begin{equation}
 \beta^2 f(r) < \beta_H^2 \ . 
\end{equation}
The pressure monotonically decreases as $r$ increases. 
The pressure must be positive at least at $r=0$ 
so that it is positive somewhere on this spacetime. 
This condition requires 
\begin{equation}
 h_0 > 1 \ . 
\end{equation}
If we use this solution to $r=r_m$, the pressure becomes negative at least at $r=r_m$. 
The pressure becomes zero at some radius $r=r_0$ in $0<r_0<r_m$. 
The fluid is trapped in the region where the pressure is positive, 
and the pressure is zero at the surface of the star. 
Thus, the solution \eqref{f}--\eqref{a=0} should be connected to the exterior solution at $r=r_0$.


\subsection{Junction with the Schwarzschild spacetime}\label{ssec:junction}

Now, we consider the junction of the solution \eqref{f}--\eqref{a=0} with the exterior solution. 
Although there would be a layer near the surface of the star 
where the solution \eqref{f}--\eqref{a=0} would not give a very good approximation, 
we just ignore this layer and assume that the solution \eqref{f}--\eqref{a=0} 
is directly connected to the exterior solution. 
We also neglect the small distribution of the winding string field 
in the exterior solution due to the quantum penetration, 
and then, the exterior solution is given by the Schwarzschild spacetime. 
Distribution of the fluid \eqref{fluid} will end at the radius $r=r_0$ 
where the pressure becomes zero. 
Thus, the solution is given by \eqref{f}--\eqref{a=0} for $r<r_0$ 
and by the Schwarzschild spacetime for $r>r_0$. 
This picture would be understood as some sort of 
the classical approximation of the Horowitz-Polchinski model 
as we discussed in the previous section. 

The Schwarzschild solution is given by 
\begin{equation}
 ds^2 = - f_0(r) dt^2 + \frac{dr^2}{f_0(r)} + r^2 d \Omega^2 \ , 
 \label{SS}
\end{equation}
where 
\begin{equation}
 f_0(r) = 1 - \frac{r_h}{r} \ , 
\end{equation}
and $r_h$ is the Schwarzschild radius. 
The first junction condition requires that the induced metric 
on the interface at $r=r_0$ must be identical on both sides of the interface. 
The condition for angular components requires that the radial coordinate $r$ must be continuous at $r=r_0$. 
The first junction condition also gives a relation between the redshift factor in two metrices as 
\begin{equation}
 f(r_0) = f_0(r_0) \ . 
 \label{junction1}
\end{equation}
We also assume that the solution \eqref{f}--\eqref{a=0} is directly connected to 
the Schwarzschild solution, and there is nothing on the interface between two geometries. 
Thus, the second junction condition requires that the surface stress-energy tensor 
on the interface must be zero. 
This condition implies that the extrinsic curvature on the interface 
must be the same in the both sides. 
The second junction condition gives the following relations between two metrices; 
\begin{align}
 f'(r_0) &= f'_0(r_0) \ , 
 \label{junction2}
 \\ 
 f(r_0) h(r_0) &= f_0(r_0) \ . 
 \label{junction3}
\end{align}

\begin{figure}
\begin{center}
\includegraphics[scale=0.75]{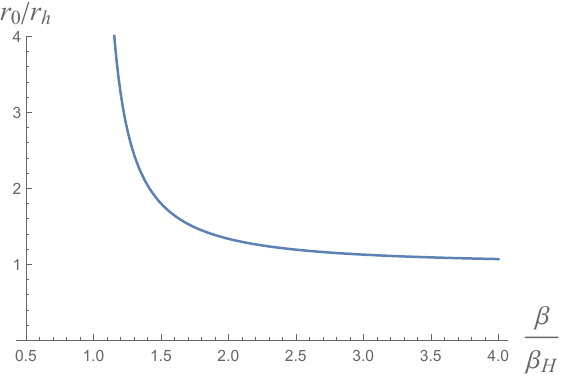}
\hspace{24pt}
\includegraphics[scale=0.75]{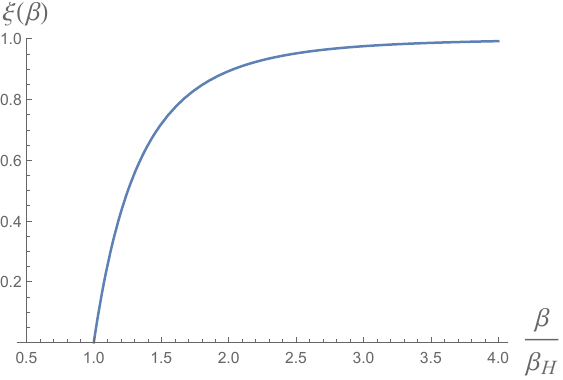}
\caption{%
The surface radius $r_0$ and the radio $\xi(\beta) = r_0^2/r_m^2$ as functions of $\beta$.  
The parameter $\xi(\beta)$ is calculated numerically. 
}\label{fig:r0}
\end{center}
\end{figure}

The interface, or equivalently, the surface of the star of winding strings 
is located at $r=r_0$ where the pressure of the fluid \eqref{fluid} becomes zero. 
The pressure becomes zero when the local temperature agrees with the Hagedorn temperature, 
\begin{equation}
 \beta^2 f(r_0) = \beta_H^2 \ . 
 \label{interface}
\end{equation}
From this condition with the junction condition \eqref{junction1}, 
we find that the position of the surface $r_0$ is determined by 
the temperature of the fluid and the Schwarzschild radius $r_h$ as 
\begin{equation}
 r_0 = \frac{\beta^2 r_h}{\beta^2 - \beta_H^2} \ . 
 \label{r0}
\end{equation}
The surface radius as a function of $\beta$ is shown in Fig.~\ref{fig:r0}(left). 

Constants $D_0$ and $h_0$ in the solution \eqref{f}--\eqref{h}, 
are determined by the junction conditions. 
The condition \eqref{interface} gives the condition 
\begin{equation}
 \frac{1}{h_0} 
 = 
 \frac{\sqrt{r_m^2 - r_0^2}}{r_0} \sin^{-1}\left(\frac{r_0}{r_m}\right) \ , 
 \label{cond1}
\end{equation}
where $r_m$ and $r_0$ are given by \eqref{rm} and \eqref{r0}, respectively. 
The junction condition \eqref{junction2} gives another condition, 
\begin{equation}
 \frac{r_h}{r_0^2} 
 = 
 \frac{\beta_H^2}{\beta^2}
 \left[ - \frac{1}{r_0}  
 + \frac{r_m^2}{r_0^2\sqrt{r_m^2 - r_0^2}} 
 \sin^{-1}\left(\frac{r_0}{r_m}\right)\right] \ . 
 \label{cond2}
\end{equation}
By using \eqref{junction1}, the condition \eqref{cond2} can be simplified as 
\begin{equation}
 \frac{1}{h_0} = \frac{\beta^2}{\beta_H^2}\left(1-\frac{r_0^2}{r_m^2}\right) \ . 
 \label{cond3}
\end{equation}
The condition \eqref{cond3} can be rewritten as $h(r_0) = 1$, 
which is equivalent to \eqref{junction3} with \eqref{junction1}. 
Thus, three junction conditions \eqref{junction1}--\eqref{junction3}
are satisfied if three parameters $D_0$, $h_0$ and $r_0$ 
are given by three conditions \eqref{r0}, \eqref{cond1} and \eqref{cond3}. 

All three parameters $D_0$, $h_0$ and $r_0$ are determined for given $\beta$ and $r_h$. 
We define the parameter $\xi$ 
in terms of the radio of $r_0$ to $r_m$ as 
\begin{equation}
 \xi(\beta) = \frac{r_0^2}{r_m^2} \ . 
 \label{xi}
\end{equation}
The value of $\xi$ is determined by the conditions \eqref{cond1} and \eqref{cond3}, namely, 
\begin{equation}
 \sqrt{ \frac{1}{\xi(\beta)} - 1}\ \sin^{-1}\left(\sqrt{\xi(\beta)}\right) 
 = 
 \frac{\beta^2}{\beta_H^2}\left(1-\xi(\beta)\right) \ , 
 \label{cond4}
\end{equation}
though this equation cannot be solved analytically. 
Since $\xi(\beta)$ is a solution of \eqref{cond4}, 
it is a function of $\beta/\beta_H$. 
Then, the integration constant $h_0$ is determined by \eqref{cond3}, 
and is expressed by using $\xi(\beta)$. 
Since $\xi(\beta)$ is a function of $\beta/\beta_H$, 
$h_0$ is also a function of $\beta/\beta_H$. 
The parameter $D_0$, which is related to normalization of $\chi$, 
is also determined by using \eqref{rm}. 

\begin{figure}
\begin{center}
\includegraphics[scale=0.75]{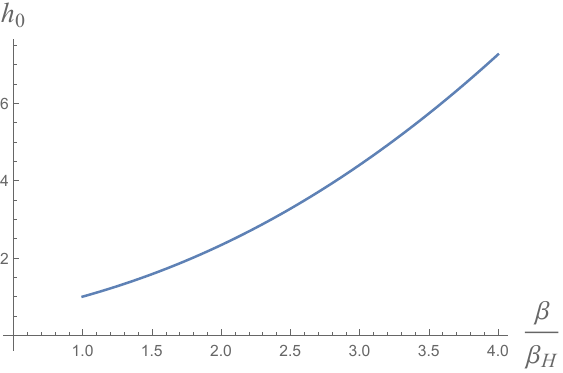}
\hspace{24pt}
\includegraphics[scale=0.75]{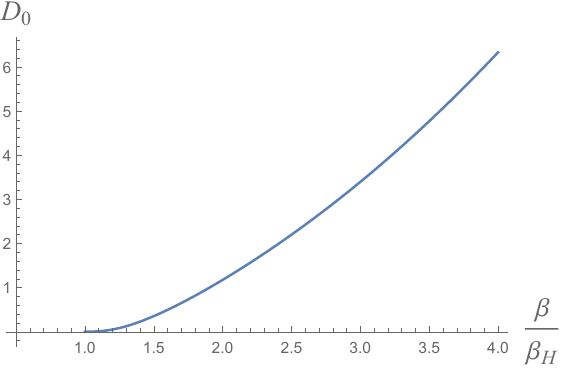}
\caption{%
Constants $h_0$ and $D_0$ by using the numerical solution of $\xi(\beta)$. 
}\label{fig:h0}
\end{center}
\end{figure}

Thus, 
constants $h_0$ and $D_0$ are calculated and are expressed as 
\begin{align}
 h_0 &= \frac{\beta_H^2}{\beta^2} \frac{1}{1-\xi(\beta)} \ , 
 \label{h0}
 \\
 D_0 &= 
 \frac{\left(\beta^2 - \beta_H^2\right)^2\xi(\beta)}{r_h^2\beta^6 \left(1-\xi(\beta)\right)} \ , 
 \label{D0}
\end{align}
and the radius of the surface $r_0$ is given by \eqref{r0}. 
Eq.~\eqref{cond4} has two solutions for $r_0 \leq r_m$. 
One solution is $r_0 = r_m$, which would be considered unphysical, 
and $\xi(\beta) = r_0/r_m$ should be determined by the other solution. 
The ratio $\xi(\beta)$ monotonically increases as the temperature $\beta_H/\beta$ decreases. 
We have 
\begin{align}
 \xi(\beta) &\to 0 \ , 
 & 
 &\text{in} \qquad \beta \to \beta_H \ , 
\\
 \xi(\beta) &\to 1 \ , 
 & 
 &\text{in} \qquad \beta \to \infty \ . 
\end{align}
Eq.~\ref{cond4} can be solved numerically, and 
the solution $\xi(\beta)$ is shown in Fig.~\ref{fig:r0}(right). 
Once we obtain $\xi(\beta)$, the constants $h_0$ and $D_0$ can also be calculated. 
The result is shown in Fig.~\ref{fig:h0}. 

Now, we calculate the ADM mass. 
Since the exterior geometry is the Schwarzschild spacetime with the Schwarzschild radius $r_h$, 
the ADM mass must be 
\begin{equation}
 M = \frac{r_h}{2 G_N} \ . 
 \label{SSmass0}
\end{equation}
We will calculate the ADM mass by using the formula \eqref{ADM0} 
for the fluid of winding strings, and see that it agrees with \eqref{SSmass0}. 
By using the energy density and pressure of the fluid \eqref{fluid}, 
the ADM mass is calculated as \eqref{ADM}. 
Substituting the solution \eqref{f}--\eqref{a=0}, 
the ADM mass becomes 
\begin{align}
 M 
 &= 
 \frac{1}{4\pi G_N}\frac{\beta_H^3 D_0}{\beta \sqrt{h_0}} 
 \int dr\,4\pi r^2 \left(1 - \frac{r^2}{r_m^2}\right)^{-1/2} 
 \\
 &=  
 \frac{1}{2 G_N}\frac{r_m^3 \beta_H^3 D_0}{\beta \sqrt{h_0}} 
 \left[- \frac{r_0}{r_m}\sqrt{1-\frac{r_0^2}{r_m^2}} + \sin^{-1}\left(\frac{r_0}{r_m}\right)\right] \ . 
\end{align}
By using \eqref{cond4} and \eqref{xi}, 
it can be rewritten as 
\begin{equation}
 M = \frac{r_0^3\beta_H (\beta^2-\beta_H^2) \sqrt{1-\xi(\beta)}\, D_0}
 {2 G_N \beta\, \xi(\beta)\sqrt{h_0}} 
 \ . 
\end{equation}
Then, substituting \eqref{h0}, \eqref{D0} and \eqref{r0}, we obtain 
\begin{align}
 M 
 &= 
 \frac{r_h}{2 G_N} \ . 
 \label{SSmass}
\end{align}
Thus, we have reproduced the consistency condition. 

The entropy of the fluid can also be calculated by using \eqref{Entropy}, 
\begin{equation}
 S 
 = 
 \frac{\beta^3 D_0}{4\pi G_N} 
 \int dr\,4\pi r^2 f(r) h^{-1/2}(r) 
 \ , 
\end{equation}
In a similar fashion to the ADM mass, 
substituting the solution \eqref{f}--\eqref{a=0} with \eqref{r0}, \eqref{h0} and \eqref{D0}, 
we obtain 
\begin{equation}
 S 
 = 
 \frac{r_h}{4 G_N}\left[3\beta + 2 \frac{\beta^3}{\beta_H^2} 
 - \frac{2\beta^5\xi(\beta)}{\beta_H^2\left(\beta^2 - \beta_H^2\right)}\right] 
 \ . 
 \label{EntropyHP}
\end{equation}


\subsection{Quantization condition}\label{ssec:quantization}

We have seen that the solution has two parameters $\beta$ and $r_h$ at the classical level. 
In this fluid model, the size of the star depends on the Schwarzschild radius. 
In the original Horowitz-Polchinski model, the solution is obtained for a given temperature. 
The size and total mass of the solution are determined by the temperature. 
Thus, in the Horowitz-Polchinski model, 
the Schwarzschild radius should also be given as a function of the temperature. 
Solving the equation for $\chi$ corresponds to imposing the quantization condition. 
Two parameters in the fluid model, $\beta$ and $r_h$, are not independent to each other 
but should be related to each other by the quantization condition. 


In the classical limit, the quantization condition is approximately given by 
\begin{equation}
 \int p_r dr \sim 2\pi n \ . 
\end{equation}
Although this quantization condition is good for large $n$, 
the size of the ground state can be estimated by taking $n=1$. 
For the ground state, the size can also be read off from 
the uncertainty principle, 
\begin{equation}
 (\Delta r)\times (\Delta p_r) = 2\pi \ , 
 \label{uncertainty}
\end{equation}
since the state is localized around a point in the phase space. 
Although we have assumed that the momentum is negligible 
in the stress-energy tensor \eqref{Ttt}--\eqref{Tyy}, 
the uncertainty of the momentum would be estimated as 
\begin{equation}
 (\Delta p_r)^2 + m_\text{eff}^2 = 0 \ . 
\end{equation}
Thus, the size of the solution is approximately given by 
\begin{align}
 r_0 \sim (\Delta r) 
 &= 
 \frac{\alpha'}{\sqrt{\beta_H^2 - f(0)\beta^2}} 
\notag\\
 &= 
 \frac{\alpha'}{\sqrt{\beta_H^2 - \beta^2 + \beta^2 \xi(\beta)}}   
 \ . 
 \label{q-naive}
\end{align}
By using \eqref{r0}, the Schwarzschild radius is also determined by the temperature of the fluid as 
\begin{equation}
 r_h 
 \sim 
 \frac{\alpha' \left(\beta^2 - \beta_H^2\right)}{\beta^2\sqrt{\beta_H^2 - \beta^2 + \beta^2 \xi(\beta)}} \ . 
\end{equation}
Note that the temperature of the fluid is different from the Hawking temperature, 
since the Schwarzschild spacetime is connected to the interior solution \eqref{f}--\eqref{a=0} 
outside the Schwarzschild radius, and the geometry has no event horizon. 

Unfortunately, this rough estimation is not very good for the low temperature limit $\beta\to\infty$. 
Here, we used the maximal radius and momentum for 
the uncertainty $\Delta r$ and $\Delta p_r$ in \eqref{uncertainty}, 
but actuall area in the phase space might be smaller than $(\Delta r)\times (\Delta p_r)$. 
Thus, this rough estimation \eqref{q-naive} just gives lower bound of the size. 
In the next section, we will study two cases of the high and low temperature limit 
and see the quantization condition in more detail. 


\section{Black hole--string transition}\label{sec:Limits}

Although we have solved the Einstein equation for the winding string fluid analytically, 
we still have not obtained intuitive descriptions of the solution 
as we do not have the analytic expression of the parameter $\xi(\beta)$. 
In order to have a better understanding of the solution, 
it is convenient to study some special cases of the solution. 

In this section, we consider the structure of the solution in high and low temperatures. 
At high temperatures near the Hagedorn temperature, 
the solution has consistent behaviors with the Horowitz-Polchinski solution, 
and hence, describes winding strings which are weakly bounded by the self-gravitation. 
In the low temperature limit, the solution approaches the Schwarzschild black hole. 
The solution has approximately the same size, mass, temperature and entropy to the Schwarzschild black hole.  
To be more precise, the size is slightly larger than the Schwarzscihld radius, 
and hence, the solution has no event horizon. 
Although we focus only on two limits of high and low temperatures here, 
our solution is parameterized by the temperature smoothly, 
and the solution at any temperature between these two limits 
can be obtained just by choosing the parameter accordingly. 
Thus, our fluid model of winding strings gives 
a description of the transition between black holes and strings. 
Since the solution at high temperature continues to 
the solution at low temperatures smoothly, 
the self-gravitating string phase and black hole phase are 
continuously connected to each other. 


\subsection{High temperature limit: a bound state of strings}\label{ssec:high}

Here, we consider the high temperature limit. 
We take the limit in which the temperature approaches the Hagedorn temperature, 
$\beta\to \beta_H$, and calculate the leading order terms in this limit. 

We first calculate $\xi(\beta)\,\left(=r_0^2/r_m^2\right)$, as 
most of parameters of the solution \eqref{f}--\eqref{a=0} are given in terms of $\xi(\beta)$. 
The parameter $\xi(\beta)$ is determined by the condition \eqref{cond4}. 
At the Hagedorn temperature, we have $\xi(\beta_H) = 0$. 
Thus, it is expected that $\xi(\beta)$ is very small 
at the linear order in the small-$(\beta - \beta_H)$ limit.  
We expand \eqref{cond4} for small $\xi(\beta)$ as 
\begin{equation}
 1 - \frac{1}{3} \xi(\beta) 
 \simeq 
 \frac{\beta^2}{\beta_H^2}\left(1-\xi(\beta)\right) \ . 
\end{equation}
Then, $\xi(\beta)$ is obtained as 
\begin{equation}
 \xi(\beta) \simeq \frac{3\left(\beta^2 - \beta_H^2\right)}{3\beta^2 - \beta_H^2} 
 \simeq \frac{3\left(\beta^2 - \beta_H^2\right)}{2\beta_H^2} 
 \ . 
\end{equation}
Substituting this solution to \eqref{h0} and \eqref{D0}, 
the constants $h_0$ and $D_0$ are calculated as 
\begin{align}
 h_0 
 &\simeq 
 \frac{\left(3\beta^2 - \beta_H^2\right)}{2\beta^2}
 \simeq 1 + \frac{\beta^2 - \beta_H^2}{2\beta_H^2} \ , 
 \label{h0-high}
 \\
 D_0 
 &\simeq 
 \frac{3\left(\beta^2 - \beta_H^2\right)^3}{2r_h^2\beta^6\beta_H^2} 
 \simeq \frac{3\left(\beta^2 - \beta_H^2\right)^3}{2r_h^2\beta_H^8} \ . 
 \label{D0-high}
\end{align}
Since the solution \eqref{f}--\eqref{a=0} describes 
spacetime in the region $r\leq r_0$, 
$r/r_m$ is also very small near the Hagedorn temperature; 
\begin{equation}
 \frac{r^2}{r_m^2} \leq \frac{r_0^2}{r_m^2} 
 \simeq 
 \frac{3\left(\beta^2 - \beta_H^2\right)}{2\beta_H^2} = \mathcal O(\beta-\beta_H) \ . 
\end{equation}
Substituting \eqref{h0-high} and \eqref{D0-high} to \eqref{rm}, 
$r_m$ is expressed as 
\begin{equation}
 r_m^2 \simeq \frac{2r_h^2\beta_H^6}{3\left(\beta^2 - \beta_H^2\right)^3} \ . 
\end{equation}
Thus, the solution 
$f(r)$ (with $a=0$) and $h(r)$ are expanded as  
\begin{align}
 f(r) 
 &= 
 1 - \frac{3(\beta^2 - \beta_H^2)}{2\beta_H^2} 
 + \frac{\left(\beta^2 - \beta_H^2\right)^3}{2r_h^2\beta_H^6}\, r^2 
 + \mathcal O\left((\beta-\beta_H)^2\right) \ , 
 \label{f-high}
 \\
 h(r) 
 &= 
 1 + \frac{3(\beta^2 - \beta_H^2)}{2\beta_H^2} 
 - \frac{3\left(\beta^2 - \beta_H^2\right)^3}{2r_h^2\beta_H^6}\, r^2 
 + \mathcal O\left((\beta-\beta_H)^2\right) \ . 
 \label{h-high}
\end{align}
At the leading order in the high temperature limit, 
the geometry can be treated as flat space $f(r) = h(r) = 1$. 
The linear order corrections are necessary 
to see the gravitational potential for the winding string field, 
or equivalently the effective mass \eqref{meff}. 
They are also necessary to calculate the curvature. 
This structure is the same to the original equations of motion of 
the Horowitz-Polchinski model \eqref{eq-phi} and \eqref{eq-chi}. 

We consider the ADM mass and the entropy of the solution. 
The ADM mass of the fluid of winding strings \eqref{fluid} 
is calculated in Sec.~\ref{ssec:junction} and is obtained as \eqref{SSmass}. 
Since the geometry is flat at the leading order in the high temperature limit, 
the ADM mass can easily be calculated directly from \eqref{ADM} as 
\begin{equation}
 M = \frac{4\pi r_0^3}{3} \frac{\beta_H^2 D_0}{4\pi G_N} \ . 
\end{equation}
By using \eqref{r0} and \eqref{D0-high}, 
the ADM mass is evaluated as 
\begin{equation}
 M = \frac{r_h}{2 G_N} \ . 
\end{equation}
This is consistent with the Schwarzschild radius in the exterior geometry. 
In a similar fashion, by using \eqref{Entropy}, the entropy is calculated as 
\begin{equation}
 S 
 = 
 \frac{4\pi r_0^3}{3} \frac{\beta^3 D_0}{4\pi G_N} 
 \simeq  
 \frac{\beta_H r_h}{2 G_N} \ . 
\end{equation}
At the leading order of in the high temperature limit, 
the ADM mass and entropy satisfy the relation,  
\begin{equation}
 S \simeq \beta M \ . 
\end{equation}
This relation can also be obtained from the local thermodynamic relation, 
since the pressure vanishes at the leading order of $\beta \to \beta_H$. 

The quantization condition is expressed as 
\begin{equation}
 \int_0^{r_0} dr \sqrt{- g_{rr} m_\text{eff}^2} = \frac{3}{4}\pi \ , 
 \label{q-WKB}
\end{equation}
where the numerical factor in r.h.s.\ is estimated from 
the junction condition of the WKB approximation at the classical turning point. 
By using \eqref{f-high} and \eqref{h-high}, 
the effective mass becomes 
\begin{equation}
 m_\text{eff}^2 
 \simeq  
 - \frac{\beta^2 - \beta_H^2}{8\pi^2 \alpha^{\prime\,2}}
 \left[1 - \frac{(\beta^2 - \beta_H^2)^2}{r_h^2\beta_H^4}\,r^2\right] \ , 
 \label{meff-high}
\end{equation}
and the radial component of the metric $g_{rr} = 1/(f(r)h(r))$ gives only higher order corrections. 
Then, the quantization condition \eqref{q-WKB} becomes 
\begin{equation}
 \frac{r_h \beta_H^2}{2^{3/2} \pi \alpha'\sqrt{\beta^2-\beta_H^2}} = 3 \ . 
\end{equation}
Thus, the Schwarzschild radius $r_h$ is given in terms of the temperature $\beta$ as 
\begin{equation}
 r_h = 2^{3/2}\cdot 3\,\pi \alpha' \frac{\sqrt{\beta^2-\beta_H^2}}{\beta_H^2} \ . 
 \label{rh-WKB}
\end{equation}
By using \eqref{r0}, the position of the surface of the star is obtained as 
\begin{equation}
 r_0 = \frac{2^{3/2}\cdot 3\,\pi \alpha'}{\sqrt{\beta^2-\beta_H^2}} \ . 
 \label{r0-high}
\end{equation}
This agrees with the typical size of the solution of Horowitz-Polchinski model \eqref{size}. 
The ADM mass and entropy are also calculated as 
\begin{align}
 M 
 &= 
 3\sqrt 2 \,\pi \alpha' \frac{\sqrt{\beta^2-\beta_H^2}}{G_N \beta_H^2}  \ , 
 &
 S 
 &= 
 3\sqrt 2 \,\pi \alpha' \frac{\sqrt{\beta^2-\beta_H^2}}{G_N \beta_H} \ . 
\end{align}

The quantization can also be estimated by solving 
the equation of motion for winding string field \eqref{eq-chi} 
by using the solution \eqref{f}--\eqref{a=0} as a background geometry. 
Near the Hagedorn temperature, the geometry is approximated by \eqref{h0-high}--\eqref{D0-high}, 
and the effective mass \eqref{meff-high} gives 
the harmonic potential $V(r) = \frac{1}{2}\omega^2 r^2$ and the eigenvalue $E$ of the ``Hamiltonian'' with 
\begin{align}
 \omega &= \frac{(\beta^2 - \beta_H^2)^{3/2}}{2^{3/2}\pi \alpha' r_h \beta_H^2} \ , 
 & 
 E &= \frac{\beta^2 - \beta_H^2}{16\pi^2 \alpha^{\prime\,2}}
\end{align}
Thus, the wave function of the ground state is given by the Gaussian function, 
\begin{equation}
 \chi \propto e^{- c r^2} \ , 
\end{equation}
where the constant $c$ is fixed by solving the equation of motion as 
\begin{equation}
 c = \frac{\beta^2 - \beta_H^2}{48\pi^2 \alpha^{\prime\,2}} \ . 
\end{equation}
Since the ground state of the three-dimensional harmonic oscillator has the energy, 
\begin{equation}
 E = \frac{3}{2} \omega \ , 
\end{equation}
the quantization condition is obtained as 
\begin{equation}
 r_h = 2^{3/2}\cdot 3\,\pi \alpha' \frac{\sqrt{\beta^2-\beta_H^2}}{\beta_H^2} \ . 
\end{equation}
This agrees with the result of the WKB approximation \eqref{rh-WKB}. 

Here, we have seen that the gravitational potential can be approximated by 
the harmonic potential by using the fluid model. 
We have solved the equation of motion for the string field $\chi$, 
to see the quantization condition. 
The solution of $\chi$ would be a better description of $\chi$, 
and can be considered as a correction to the fluid approximation. 
We can further substitute the solution of $\chi$ into the Einstein equation 
and derive a better solution of the metric, in principle. 
By repeating this procedure several times, 
we will be able to obtain a more accurate solution of the Horowitz-Polchinski model. 
However, in this paper, we will focus on the fluid model 
and do not pursue this direction.


\subsection{Low temperature limit: an approximate black hole}\label{ssec:low}

Here, we consider the low temperature limit, $\beta\to\infty$. 
First, we consider the size of the solution, $r_0$. 
The condition \eqref{r0} implies that the radius of the surface of the star 
approaches the Schwarzschild radius in the low temperature limit, namely, 
\begin{align}
 r_0 &\to r_h \ , 
 & 
 \text{in}&\qquad\qquad \beta\to\infty \ . 
\end{align}

Next, we consider the parameter $\xi(\beta)$. 
The condition \eqref{cond4} implies $\xi(\beta) \to 1$ in $\beta\to\infty$. 
Since $\xi(\beta)$ is defined as \eqref{xi}, 
the condition $\xi(\infty) = 1$ means 
\begin{equation}
 r_m \simeq r_0 \ . 
\end{equation}
In contrast to the high temperature limit, 
entire expressions of the solution \eqref{f}--\eqref{h} 
are necessary to describe the geometry in the low energy limit. 
In order to calculate the constants $h_0$ and $D_0$, 
the next-to-leading order correction of $\xi(\beta)$ is needed. 
We expand the condition \eqref{cond4} around $\xi = 1$ to obtain 
\begin{equation}
 \frac{\pi}{2}\sqrt{1-\xi(\beta)}  
 \simeq 
 \frac{\beta^2}{\beta_H^2}\left(1-\xi(\beta)\right) \ . 
\end{equation}
Then, the parameter $\xi(\beta)$ is calculated approximately as 
\begin{equation}
 \xi(\beta) = \frac{r_0^2}{r_m^2} \simeq 1 - \frac{\pi^2\beta_H^4}{4\beta^4} \ . 
\label{xi-low}
\end{equation}
By using this solution, the constants $h_0$ and $D_0$ are expressed as 
\begin{align}
 h_0 
 &\simeq 
 \frac{4\beta^2}{\pi^2\beta_H^2} \ , 
 \label{h0-low}
 \\
 D_0
 &\simeq 
 \frac{4\beta^2}{\pi^2 r_h^2\beta_H^4} \ . 
 \label{D0-low}
\end{align}

At the leading order of the large-$\beta$ expansion, 
$f(r)$ and $h(r)$ are approximated as 
\begin{align}
 f(r) 
 &= 
 \frac{\beta_H^2}{\beta^2}\left[
 1 
 - \frac{\sqrt{r_h^2 - r^2}}{r} 
 \sin^{-1}\left(\frac{r}{r_h}\right)\right] + \mathcal O(\beta^{-4}) \ , 
 \label{f-low}
 \\
 h(r) 
 &= 
 \frac{4 \beta^4}{\pi^2 \beta_H^4} \left(1-\frac{r^2}{r_h^2}\right) + \mathcal O(\beta^2) \ . 
 \label{h-low}
\end{align}
Thus, in $\beta\to\infty$, we have $f(r)\to 0$.  
From the viewpoint of the fiducial observer in the asymptotic region, $r\to\infty$, 
the redshift factor inside the star is almost zero. 
This is also obvious from size of the star --- 
the surface radius $r_0$ is approximately the same to but slightly larger than the Schwarzschild radius. 
Thus, from the viewpoint of the fiducial observer, the star approximately behaves as a black hole. 
From the viewpoint of the observer at the surface of the star $r=r_0$, 
the redshift factor at the center $r=0$ is still much smaller than the factor at the surface. 
At the center of the star, $r=0$, \eqref{f-low} gives $f(r) = \mathcal O(\beta^{-4})$, 
and hence, we need to take the higher order corrections into consideration. 
The redshift factor is expanded around $r=0$ as 
\begin{equation}
 f(r) 
 \simeq 
 \frac{\pi^2\beta_H^4}{4 \beta^4} + \frac{\beta_H^2 r^2}{3 \beta^2 r_h^2} + \cdots \ , 
\end{equation}
and hence, behaves as 
\begin{align}
 f(r) 
 &\sim 
 \mathcal O(\beta^{-4}) \ , 
 & 
 \text{for}&\qquad\qquad 
 r \ll \frac{\beta_H}{\beta} r_h \ , 
\\
 f(r) 
 &\sim 
 \mathcal O(\beta^{-2}) \ , 
 & 
 \text{for}&\qquad\qquad 
 r \gg \frac{\beta_H}{\beta} r_h \ . 
\end{align}
In a similar fashion, the radial component of the metric also behaves as 
\begin{align}
 g_{rr} &= \frac{1}{f(r)h(r)} 
 \simeq 
 1 - \frac{4 \beta^2 r^2}{3\pi^2 \beta_H^2 r_h^2} + \cdots \ , 
 & 
 &\text{for}\qquad\qquad 
 r \ll \frac{\beta_H}{\beta} r_h \ , 
\\
 g_{rr} 
 &\simeq 
 \frac{\beta^2}{\beta_H^2} + \cdots \ , 
 & 
 &\text{for}\qquad\qquad 
 r_0 - r \ll \frac{\pi^2 \beta_H^4}{8\beta^4}r_h \ , 
\\
 g_{rr} 
 &\sim 
 \mathcal O(\beta^{-2}) \ , 
 & 
 &\text{otherwise.}
\end{align}
The proper length inside the fluid is mostly much shorter than 
the (areal) radius at the surface $r_0$. 
Only very near the surface, the proper distance is much longer, 
as is near the horizon of the Schwarzschild black hole. 
At $r=0$, we have $g_{rr} = 1$, which implies no conical singularity at the origin. 

Now, we consider the quantization condition \eqref{q-WKB}. 
Here, we only make a rough order estimation because 
\eqref{q-WKB} itself is not a precise condition 
despite of its complicated expression in the low temperature limit. 
At the leading order in the large-$\beta$ limit, 
the effective mass $m_\text{eff}^2$ becomes  
\begin{equation}
 m^2_\text{eff} 
 = 
 \frac{\beta^2 f(r) - \beta_H^2}{(2\pi \alpha')^2} 
 = 
 - \frac{\beta_H^2}{(2\pi \alpha')^2} 
 \frac{\sqrt{r_h^2 - r^2}}{r} 
 \sin^{-1}\left(\frac{r}{r_h}\right)
 + \mathcal O(\beta^{-2}) \ . 
\end{equation}
It has minimum at $r=0$, and hence, we have 
\begin{equation}
 - m_\text{eff}^2 \lesssim - m_\text{eff}^2(r=0) = \frac{\beta_H^2}{(2 \pi \alpha')^2} \ . 
\end{equation}
The radial component of the metric is estimated up to $\mathcal O(1)$ factor as 
\begin{equation}
 g_{rr} \sim \frac{\beta_H^2}{\beta^2} \ . 
\end{equation}
Then, the quantization condition is roughly expressed as 
\begin{equation}
 \int_0^{r_h} \sqrt{-g_{rr}m_\text{eff}^2}\ dr 
 \sim \frac{\beta_H}{\beta} \times \frac{\beta_H}{2\pi \alpha'} \times r_h \ . 
 \label{q-low0}
\end{equation}
However, this integration gives logarithmic divergence%
\footnote{%
It can be seen from the divergence $\propto 1/r$ in the integrand. 
The integrand has another divergence $\propto (r-r_0)^{-1/2}$, 
which does not give the divergence of the integral. 
} 
if we use the expression of $f(r)$ at the leading order, \eqref{f-low}. 
This is because $\mathcal O(\beta^{-4})$ corrections should be 
taken into account near the center of the star, $r\simeq 0$.
Since $g_{rr}$ behaves as $g_{rr}\sim \mathcal O(1)$ 
only very small region $r < \frac{\beta_H}{\beta} r_h$, 
the contribution from the integration around $r=0$ is limited to 
\begin{equation}
 \int_{r< \frac{\beta_H}{\beta}r_h} \sqrt{-g_{rr}m_\text{eff}^2}\ dr 
 \sim \frac{\beta_H}{2\pi \alpha'} \times \frac{\beta_H}{\beta} r_h \ , 
\end{equation}
which is of the same order to \eqref{q-low0}. 
Thus, the quantization condition in the low temperature limit is estimated 
up to the numerical factor as 
\begin{equation}
 \frac{\beta_H^2 r_h}{2\pi \alpha' \beta} \sim 1 \ . 
  \label{q-low}
\end{equation}
From this quantization condition, 
the size of the solution, which is also approximately the same to the Schwarzschild radius, 
is estimated as 
\begin{equation}
 r_0 \simeq r_h \sim \frac{\alpha' \beta}{\beta_H^2} \ . 
\end{equation}
Since the Hagedorn temperature is the same to the string scale 
(up to the numerical factor), $\beta_H^2 \sim \alpha'$, 
the expression above implies that the relation between the radius and the temperature 
is of the same order to the Schwarzschild black hole. 
Although our rough estimation does not reproduce the numerical coefficient of the temperature, 
the solution is expected to behave approximately 
as the Schwarzschild black hole in the low temperature limit. 

We consider the ADM mass and the entropy. 
As we studied in Sec.~\ref{ssec:junction}, 
the ADM mass is given in terms of the Schwarzschild radius as \eqref{SSmass}. 
The entropy is given by \eqref{EntropyHP}. 
Since $\xi(\beta)$ behaves as \eqref{xi-low} in the low temperature limit $\beta\to\infty$, 
the entropy \eqref{EntropyHP} becomes 
\begin{equation}
 S \simeq \frac{\beta r_h}{4 G_N} \ . 
\end{equation}
By using the quantization condition \eqref{q-low}, 
the entropy is approximately proportional to the area of the surface of the star, 
which is approximately the same to the area of the horizon of the Schwarzschild black hole; 
\begin{equation}
 S \sim \frac{r_h^2}{G_N} \ . 
\end{equation}
Thus, in the low temperature limit, our solution reproduces the Bekenstein-Hawking entropy 
up to the numerical coefficient.

\section{Horizonless geometry and negative energy}\label{sec:Buchdahl}

In the previous section, we have seen that 
our fluid model of winding strings describes the transition 
between black holes and strings. 
In low temperatures, we obtained a horizonless geometry 
which approximately behaves as a black hole. 
Our result would imply either that 
black holes do not exactly have the event horizon, 
or that the string bound state turns into a black hole 
when the difference between the Schwarzschild radius and 
the size of the star becomes less than the cut-off scale. 

Here, we discuss the possibility of an approximate black hole without the exact event horizon. 
It is sometimes considered that such geometries cannot be realized 
because of the Buchdahl theorem \cite{Buchdahl:1959zz}. 
The Buchdahl theorem states that the size of the star 
must be larger than 9/8 of the Schwarzschild radius 
for static and spherically symmetric solutions 
if the star consists of a perfect fluid and 
the energy density is non-increasing outwards ($\partial_r\rho\leq 0$) and non-negative ($\rho\geq 0$). 
However, in our solution, the size of the star of winding strings 
can be arbitrarily close to the Schwarzschild radius by taking a sufficiently low temperature, 
and hence, the solution does not satisfy the Buchdahl inequality. 
The inequality is violated because our solution does not satisfy 
an assumption of the Buchdahl theorem --- 
the energy density is increasing outwards and can even be negative around the center of the star. 

The energy density $\rho$ of winding strings behaves as 
\begin{equation}
 \rho = \frac{3\beta^2 f(r) - \beta_H^2}{16\pi G_N \left(2\pi \alpha'\right)^2} \ , 
\end{equation}
and $f(r)$ is a monotonic function which has minimum at $r=0$. 
Thus, in contrast to ordinary fluids, the fluid of winding strings 
has a larger energy density in outer places. 
In fact, it is straightforward to see that 
\begin{equation}
 \partial_r \rho = \frac{3\beta^2 f'(r)}{16\pi G_N \left(2\pi \alpha'\right)^2} > 0 \ , 
\end{equation}
for $0<r<r_0$ by using our solution \eqref{f}--\eqref{a=0}. 
Hence, the solution can violate the Buchdahl inequality. 

Moreover, the energy density $\rho$ becomes negative at sufficiently low temperatures. 
The energy density at $r=0$ is given by 
\begin{equation}
 \rho(r=0) = \frac{\beta_H^2 D_0}{8\pi G_N} \left(\frac{3}{h_0} - 1\right) \ . 
\end{equation}
Thus, the energy density becomes negative around $r=0$ if 
\begin{equation}
 h_0 \geq 3 \ , 
\end{equation}
or equivalently, at temperatures below some critical temperature $\beta_c$. 
The critical temperature is estimated from the low energy limit as 
\begin{equation}
 \frac{\beta_c}{\beta_H} = \frac{\sqrt 3 \pi}{2} \simeq 2.72 \ , 
\end{equation}
from the high temperature limit as 
\begin{equation}
 \frac{\beta_c}{\beta_H} = \sqrt{5} \simeq 2.24 \ , 
\end{equation}
or numerically estimated as 
\begin{equation}
 \frac{\beta_c}{\beta_H} \simeq 2.37 \ . 
\end{equation}

Although the negative energy density is unnatural feature for ordinary fluids, 
it is not surprising that the fluid of winding strings has the negative energy density. 
The origin of this negative energy density is 
the tachyonic mass of winding strings at temperatures beyond the Hagedorn temperature. 
The energy density can be separated as 
\begin{equation}
 \rho 
 = 
 \frac{2\beta^2 f(r)}{16\pi G_N (2\pi \alpha')^2}\left|\chi\right|^2 
 + \frac{m_\text{eff}^2}{16\pi G_N} \left|\chi\right|^2 \ , 
\end{equation}
where the first term would be interpreted as the thermal energy 
and the second term is the potential energy which comes from the effective mass. 
%
The fluid is localized in the region where the pressure is positive, 
\begin{equation}
 0 < P 
 = \frac{\beta_H^2 - \beta^2 f(r)}{16\pi G_N \left(2\pi \alpha'\right)^2} 
 = \frac{- m_\text{eff}^2}{16\pi G_N} \ , 
\end{equation}
implying that the local temperature of the winding string fluid 
always exceeds the Hagedorn temperature due to the blue-shift, 
although the temperature at spatial infinity is lower than the Hagedorn temperature. 
The square of the effective mass is negative, 
and winding strings become tachyonic. 
The existence of a tachyonic field implies the instability 
which leads to the decay into the true vacuum. 
The solution of the Horowitz-Polchinski model can be understood as 
the result of the tachyon condensation, and in fact, 
the winding string field $\chi$ becomes non-zero around the region where the mass is tachyonic.%
\footnote{%
To be more precise, the solution is not in the bottom of the potential 
but stabilized by the connection to the flat spacetime in the spatial infinity. 
In flat spacetime, the true vacuum is at $\chi=0$ as the mass of the winding string field is positive. 
The winding string field $\chi$ in the winding condensate cannot be very different from 
$\chi=0$ in the spatial infinity to satisfy the equation of motion, and hence, is finite.  
} 
Thus, the potential energy which comes from the effective mass is negative 
inside the bound state of winding strings.%
\footnote{%
In general, the energy in the tachyonic vacuum can be non-zero, and 
the true vacuum may not have negative energy. 
In the case of the Horowitz-Polchinski model, the energy at $\chi=0$ must be zero 
so that the energy in flat spacetime outside the bound state is zero, 
and hence, the winding condensate $\chi\neq 0$ has negative potential energy. 
} 
If the temperature is sufficiently low, the negative energy of the winding condensate 
overcomes the thermal energy, and the energy density of the fluid also becomes negative. 
In contrast to the energy density, the ``density'' of the ADM mass, 
namely the integrand of \eqref{ADM} is always positive. 
The additional term in the ADM mass is often interpreted 
as the energy from the gravitation. 
Thus, the energy of the fluid including gravitational effects is always positive.

Although the winding string field has the tachyonic mass, 
the winding condensate should be distinguished from the ordinary tachyon condensation. 
Winding strings become tachyonic by wrapping the Euclidean time circle. 
Before the Wick rotation, the Lorentzian time is not compactified, 
and hence, winding strings do not exist as itself but describe 
strings created by the thermal energy. 
In order to create massive strings,%
\footnote{%
In the worldsheet picture, the lowest mode of winding strings wrapping on the Euclidean time circle 
and massless strings propagating in the time direction correspond to 
different limits in the moduli space, respectively. 
Thus, the low energy effective theory of winding strings does not give a good description 
of massless strings but includes massive strings in the Lorentzian picture. 
} 
the temperature must be higher than the string scale, 
and hence, the winding condensate appears only in the region 
where the local temperature exceeds the Hagedorn temperature. 
%
%

When the temperature exceeds the Hagedorn temperature, 
the mass of the winding string field becomes tachyonic, implying an instability.  
Oscillations of strings give an exponentially large number of particle species, and hence,  
the entropy of highly excited strings becomes 
\begin{equation}
 S(E) \simeq \beta_H E \ . 
 \label{hes}
\end{equation}
If the system is in contact with a heat bath with a temperature higher than the Hagedorn temperature, 
highly excited strings continue to absorb the energy from the heat bath without raising the temperature. 
More and more strings are created indefinitely, and hence, 
the system becomes unstable beyond the Hagedorn temperature. 

%
The free string picture above is valid only around 
the tachyonic vacuum $\chi= 0$ of the winding string field. 
After sufficient amount of massive strings are created, 
the system possibly reaches the true vacuum. 
In the case of bound states of strings, 
the solution of the Horowitz-Polchinski model can be interpreted as the true vacuum. 
The local temperature inside the bound state exceeds the Hagedorn temperature, 
but the temperature at spatial infinity, which can be interpreted as 
the temperature of the heat bath, is lower than the Hagedorn temperature.%
\footnote{%
As it can be seen in \eqref{size} or \eqref{r0-high}, 
the size of the bound state becomes infinitely large 
as the temperature approaches the Hagedorn temperature, 
implying that winding strings cannot form a bound state 
if the temperature at spatial infinity exceeds the Hagedorn temperature. 
} 
Thus, strings inside the bound state cannot absorb the thermal energy directly from the heat bath. 
Strings inside the bound state would behave as a fluid approximately, 
and shows different thermal behavior from the free strings. 


The fluid of winding strings obeys the same thermodynamic relation 
to \eqref{hes} at the Hagedorn temperature, 
\begin{equation}
 s = \beta_H \left(\rho + P\right) = \beta_H \rho \ , 
\end{equation}
but has different behavior above the Hagedorn temperature.%
\footnote{%
When the local temperature is lower than the Hagedorn temperature, 
the pressure \eqref{fluid} becomes negative implying that 
the fluid cannot appear inside the bound state. 
} 
The local temperature of the fluid can be higher than the Hagedorn temperature, 
and the entropy and energy density \eqref{fluid} decreases as the local temperature is raised. 
This is quite different from ordinary fluids but 
reminiscent the thermodynamic property of black holes. 
This behavior also plays an important role in the violation of the Buchdahl inequality. 
The local temperature is higher in inner places due to the blue-shift 
as gravity is stronger there. 
For ordinary fluids, the energy density will be higher at higher temperatures, 
and hence, it is reasonable to assume that the energy density is non-increasing outwards. 
However, the winding string fluid has smaller energy density at higher local temperatures, 
and hence, the energy density increases outwards. 
As the Buchdahl inequality can be violated if the energy density increases outwards, 
a star of the winding string fluid can have a size smaller than 9/8 of the Schwarzschild radius. 

Although the energy density of the winding string fluid is smaller at higher temperatures, 
the density of ADM mass \eqref{ADM}, or equivalently, the energy density 
including the gravitational energy is independent of the temperature for $D=4$ 
and is larger at higher temperature for $D>4$. 
This implies that the energy of the winding strings would be 
converted into the gravitational energy, and hence, 
the energy density of winding strings decreases as the temperature is raised. 
Thus, gravitational effects would play an important role for 
the peculiar thermal behavior of the winding string fluid.

\section{Higher dimensions}\label{sec:Higher}

In this section, we consider the higher dimensional generalization of 
the solution in Sec.~\ref{ssec:solution}. 
Calculations can be proceeded in a similar fashion to previous sections, 
and hence, we do not explain the details and basically describe only the results, here. 

We first solve the Einstein equation \eqref{Ein} for $D>4$. 
The metric is given in the same form to \eqref{metric}, 
but now the 2-sphere is replaced by $(D-2)$-sphere. 
The fluid equation gives the same condition that $\chi$ is a constant, 
and we define $D_0$ by \eqref{D0-def}, again. 

By virtue of the spherical symmetry, the solution is obtained by 
solving only two components of the Einstein equation, as in the case of four dimensions. 
We consider $(t,t)$- and $(r,r)$-components, which can be expressed as 
\begin{align}
 0 &= 
 (D-2) r \left(f(r)h(r)\right)' + (D-2)(D-3) f(r) h(r) 
 \notag\\
 &\quad
 + 2 r^2 \left(3\beta^2 f(r) - \beta_H^2\right) D_0 - (D-2)(D-3) \ , 
 \\
 0 &= 
 (D-2) r f'(r) h(r) + (D-2)(D-3) f(r) h(r) 
 \notag\\
 &\quad
 + 2 r^2 \left(\beta^2 f(r) - \beta_H^2\right) D_0 - (D-2)(D-3) \ . 
\end{align}
It is straightforward to solve these equations to obtain 
\begin{align}
 f(r) 
 &= 
 \frac{\beta_H^2}{\beta^2} 
 \biggl[
  \frac{1}{h_0} {}_2F_1\left(1,\frac{D-4}{2},\frac{D-1}{2},\frac{r^2}{r_m^2}\right)
  + \frac{1}{D-1} \frac{r^2}{r_m^2} {}_2F_1\left(1,\frac{D-2}{2},\frac{D+1}{2},\frac{r^2}{r_m^2}\right)
 \biggr]
 \notag\\
 &\quad
 + a r^{3-D} \sqrt{1-\frac{r^2}{r_m^2}} \ , 
\\
 h(r) 
 &= 
 \frac{\beta^2}{\beta_H^2} h_0 \left(1 - \frac{r^2}{r_m^2}\right) \ ,  
\end{align}
where ${}_2F_1(a,b,x,x)$ is the hypergeometric function, 
$a$ and $h_0$ are the integration constants, and $r_m$ is now defined by 
\begin{equation}
 r_m^2 = \frac{(D-2)h_0}{2\beta_H^2 D_0} \ . \label{rm-higher}
\end{equation}
In order to remove the singularity at $r=0$, we take 
\begin{equation}
 a=0 \ . 
\end{equation}
The solution is defined in $0\leq r \leq r_m$, 
and $f(r)$ is a monotonic function with 
\begin{align}
 f(0) &= \frac{\beta_H^2}{\beta^2} \frac{1}{h_0} \ , 
 &
 f(r_m) &= \frac{\beta_H^2}{\beta^2} \left(1 + \frac{D-3}{h_0}\right) \ . 
\end{align}

Here, we show some examples of the solution of $f(r)$. 
\begin{itemize}
\item For $D=5$, 
\begin{equation}
 f(r) 
 = 
 \frac{\beta_H^2}{\beta^2}
 \left[-1 + 2 \left(1 + \frac{1}{h_0}\right)\frac{r_m^2}{r^2}\left(1 - \sqrt{1-\frac{r^2}{r_m^2}}\right)\right] \ . 
\end{equation}

\item
For $D=6$, 
\begin{equation}
 f(r) 
 = 
 \frac{\beta_H^2}{2\beta^2} 
 \left\{
 -1 + 3 \left(1 + \frac{2}{h_0}\right)
 \frac{r_m^2}{r^2}\left[1 - \frac{\sqrt{r_m^2 - r^2}}{r} \sin^{-1}\left(\frac{r}{r_m}\right)\right]
 \right\} \ . 
\end{equation}

\item
For $D=7$, 
\begin{equation}
 f(r) 
 = 
 \frac{\beta_H^2}{3\beta^2} 
 \left\{ 
 - 1
 + 8 \left(1 + \frac{3}{h_0}\right) \frac{r_m^4}{r^4} 
 \left[
 \left(1 - \frac{r^2}{2r_m^2}\right) 
 - \sqrt{1-\frac{r^2}{r_m^2}}
 \right]
 \right\} \ . 
\end{equation}

\end{itemize}

Next, we consider the junction condition to the Schwarzschild spacetime. 
The higher-dimensional Schwarzschild metric is given by \eqref{SS}, 
but now $f_0(r)$ is given by 
\begin{equation}
 f_0(r) = 1 - \frac{r_h^{D-3}}{r^{D-3}} \ . 
\end{equation}
By using \eqref{interface} with the junction condition \eqref{junction1}, 
the size of the star is related to the Schwarzschild radius as 
\begin{equation}
 r_0 = \left(\frac{\beta^2 - \beta_H^2}{\beta^2}\right)^\frac{1}{3-D} r_h \ . 
\end{equation}
We define $\xi(\beta)$ by the same relation \eqref{xi}. 
The parameter $\xi(\beta)$ is the solution of the equation 
\begin{equation}
 1 - \frac{\xi(\beta)}{D-1} {}_2F_1\left(1,\frac{D-2}{2},\frac{D+1}{2},\xi(\beta)\right) 
 = 
 \frac{\beta^2}{\beta_H^2} \left(1-\xi(\beta)\right) 
 {}_2F_1\left(1,\frac{D}{2},\frac{D-1}{2},\xi(\beta)\right) \ , \label{cond-higher}
\end{equation}
which comes from the junction condition for the radio $r_0/r_m$. 
By using the junction condition \eqref{junction1}--\eqref{junction3}, 
the integration constant $h_0$ is expressed in terms of $\xi(\beta)$ as 
\begin{equation}
 h_0 = \frac{\beta_H^2}{\beta^2} \frac{1}{1-\xi(\beta)} \ . 
\end{equation}
which is the same expression to the case of $D=4$. 
The constant $D_0$ is related to $h_0$ and $\xi(\beta)$ 
by \eqref{xi} and \eqref{rm-higher} and expressed as 
\begin{equation}
 D_0 
 = 
 \frac{(D-2)\left(\beta^2 - \beta_H^2\right)^{\frac{2}{D-3}} \xi(\beta)}
 {2 r_h^2\,\beta^{ \frac{2(D-1)}{D-3}} \left(1-\xi(\beta)\right)} \ . 
\end{equation}

Now, we consider the high temperature limit. 
As we discussed in the case of $D=4$, 
$\xi(\beta)$ approaches zero in the high temperature limit. 
Then, the condition \eqref{cond-higher} can be expanded as 
\begin{equation}
 \frac{\beta_H^2}{\beta^2} = 1 - \frac{\xi(\beta)}{D-1} + \mathcal O(\xi^2) \ . 
\end{equation}
Then, $\xi(\beta)$ is obtained as 
\begin{equation}
 \xi(\beta) = \frac{(D-1)\left(\beta^2 - \beta_H^2\right)}{2 \beta_H^2} 
 + \mathcal O\left((\beta^2-\beta_H^2)^2\right) \ . 
\end{equation}
Parameters of the solution, $h_0$, $D_0$ and $r_m$ are expanded as 
\begin{align}
 h_0 
 &= 
 1 - \frac{(D-3)\left(\beta^2 - \beta_H^2\right)}{2 \beta_H^2}
 + \mathcal O\left((\beta^2 - \beta_H^2)^2\right) \ , 
 \\
 D_0 
 &= 
 \frac{(D-1)(D-2)\left(\beta^2 - \beta_H^2\right)^{ \frac{D-1}{D-3}}}
 {4 r_h^2 \beta_H^{ \frac{4(D-2)}{D-3}}} 
 + \mathcal O\left((\beta^2 - \beta_H^2)^{\frac{2(D-2)}{D-3}}\right) \ , 
 \\
 r_m^2 
 &= 
 \frac{2 r_h^2 \beta_H^{\frac{2(D-1)}{D-3}}} 
 {(D-1)\left(\beta^2 - \beta_H^2\right)^{\frac{D-1}{D-3}}} 
 + \mathcal O\left((\beta^2 - \beta_H^2)^{-\frac{2}{D-3}}\right) \ . 
\end{align}
The solution of $f(r)$ and $h(r)$ are expanded as 
\begin{align}
 f(r) 
 &= 
 1 - \frac{(D-1)\left(\beta^2 - \beta_H^2\right)}{2\beta_H^2} 
 + \frac{(D-3)\left(\beta^2 - \beta_H^2\right)^{\frac{D-1}{D-3}}}
 {2 r_h^2 \beta_H^{\frac{2(D-1)}{D-3}}} r^2 
 + \mathcal O\left((\beta^2 - \beta_H^2)^{2}\right) \ , 
 \\
 h(r) 
 &= 
 1 + \frac{(D-1)\left(\beta^2 - \beta_H^2\right)}{2\beta_H^2} 
 - \frac{(D-1)\left(\beta^2 - \beta_H^2\right)^{\frac{D-1}{D-3}}}
 {2 r_h^2 \beta_H^{\frac{2(D-1)}{D-3}}} r^2 
 + \mathcal O\left((\beta^2 - \beta_H^2)^{2}\right) \ . 
\end{align}

Now, we calculate the ADM mass and entropy. 
The ADM mass is obtained as 
\begin{equation}
 M 
 \simeq 
 \frac{\beta_H^2 r_0^{D-1} D_0 \Omega_{D-2}}{4\pi G_N} 
 \simeq 
 \frac{(D-2) r_h^{D-3} \Omega_{D-2}}{16\pi G_N} \ , 
\end{equation}
where $\Omega_{D-2}$ is the area of the unit $(D-2)$-sphere. 
In a similar fashion, the entropy is calculated as 
\begin{equation}
 S 
 \simeq 
 \frac{\beta^3 r_0^{D-1} D_0 \Omega_{D-2}}{4\pi G_N} 
 \simeq 
 \frac{(D-2) \beta r_h^{D-3}\Omega_{D-2}}{16\pi G_N} \ . 
\end{equation}

The effective mass is expanded as 
\begin{equation}
 m_\text{eff}^2 
 = 
 - \frac{(D-3)\left(\beta^2 - \beta_H^2\right)}{32\pi^2\alpha^{\prime\,2}} 
 \left[
 1 - \frac{\left(\beta^2 - \beta_H^2\right)^{\frac{2}{D-3}}}{r_h^2 \beta_H^{\frac{2}{D-3}}} r^2 
 \right] 
 + \mathcal O\left((\beta^2 - \beta_H^2)^{2}\right)  
 \ . 
\end{equation}
Then, the quantization condition becomes 
\begin{equation}
 \frac{(D-3)\left(\beta^2 - \beta_H^2\right)}{8\pi^2\alpha^{\prime\,2}} 
 \simeq  
 \frac{(D-1)^2}{4}\frac{\left(\beta^2 - \beta_H^2\right)^{\frac{2}{D-3}}}{r_h^2 \beta^{\frac{2}{D-3}}} \ . 
\end{equation}
Thus, the Schwarzschild radius $r_h$ is related to the temperature of the fluid as 
\begin{equation}
 r_h 
 \simeq  
 \frac{2^{3/2}\pi \alpha' (D-1) \left(\beta^2 - \beta_H^2\right)^{-\frac{D-5}{2(D-3)}}}
 {\sqrt{D-3}\,\beta_H^{\frac{2}{D-3}}} \ . 
\end{equation}
The radius of the surface of the star agrees with the rough estimation of the scale of the solution; 
\begin{equation}
 r_0 
 = 
 \frac{2^{3/2}\pi \alpha' (D-1)}{\sqrt{(D-3)\left(\beta^2 - \beta_H^2\right)}} \ . 
\end{equation}
The ADM mass and entropy become 
\begin{align}
 M 
 &\simeq 
 \frac{(2\pi)^{\frac{3(D-3)}{2}} \alpha^{\prime\,D-3} (D-2)(D-1)^{\frac{D-1}{2}}}
 {(D-3)^{\frac{D-1}{2}}\Gamma\left(\frac{D+1}{2}\right) G_N
 \beta_H^2\left(\beta^2 - \beta_H^2\right)^{\frac{D-5}{2}}} \ , 
 \\
 S 
 &\simeq 
 \frac{(2\pi)^{\frac{3(D-3)}{2}} \alpha^{\prime\,D-3} (D-2)(D-1)^{\frac{D-1}{2}}}
 {(D-3)^{\frac{D-1}{2}}\Gamma\left(\frac{D+1}{2}\right) G_N
 \beta_H\left(\beta^2 - \beta_H^2\right)^{\frac{D-5}{2}}} \ .   
\end{align}

Next, we consider the low temperature limit. 
In the low temperature limit $\beta\to\infty$, we have $\xi(\beta)\to 1$, 
and hence, we expand \eqref{cond-higher} around $\xi(\beta) = 1$. 
Then, we obtain 
\begin{equation}
 2\sqrt{\pi}\,\Gamma\left(\frac{D+1}{2}\right) \frac{\beta_H^2}{\beta^2}
 = 
 (D-1)(D-3)\, \Gamma\left(\frac{D-2}{2}\right) \sqrt{1-\xi(\beta)} \ . 
\end{equation}
Then, $\xi(\beta)$ is solved as 
\begin{equation}
 \xi(\beta) 
 \simeq 
 1 - \frac{\pi\,\Gamma^2\!\left(\frac{D-3}{2}\right) \beta_H^4}{4\,\Gamma^2\!\left(\frac{D-2}{2}\right) \beta^4} \ . 
\end{equation}
Then, the parameters of the solution, $h_0$, $D_0$ and $r_m$ 
are approximated at the leading order as  
\begin{align}
 h_0 
 &\simeq 
 \frac{4\,\Gamma^2\!\left(\frac{D-2}{2}\right) \beta^2}
 {\pi\,\Gamma^2\!\left(\frac{D-3}{2}\right) \beta_H^2} \ , \label{h0-higher-low}
\\
 D_0 
 &\simeq 
 \frac{2(D-2)\,\Gamma^2\!\left(\frac{D-2}{2}\right) \beta^2}
 {\pi\,\Gamma^2\!\left(\frac{D-3}{2}\right) r_h^2 \beta_H^4} \ , \label{D0-higher-low}
\\
 r_m &\simeq r_0 \simeq r_h \ . \label{r0-higher-low}
\end{align}
At the leading order of the low temperature limit, 
$f(r)$ and $h(r)$ become 
\begin{align}
 f(r) 
 &= 
 \frac{1}{D-1} \frac{\beta_H^2}{\beta^2} \frac{r^2}{r_h^2}\, 
 {}_2F_1\left(1,\frac{D-2}{2},\frac{D+1}{2},\frac{r^2}{r_h^2}\right)
 \ , 
\\
 h(r) 
 &= 
 \frac{4\,\Gamma^2\!\left(\frac{D-2}{2}\right) \beta^4}
 {\pi\,\Gamma^2\!\left(\frac{D-3}{2}\right) \beta_H^4} 
 \left(1 - \frac{r^2}{r_h^2}\right) \ ,  
\end{align}

The quantization condition can be estimated in a similar fashion to the case of $D=4$. 
At the leading order of the low temperature limit, 
the effective mass is approximated as 
\begin{equation}
 m_\text{eff} 
 = 
 - \frac{\beta_H^2}{(2\pi \alpha')^2} 
 \left[
 1 - \frac{1}{D-1} \frac{\beta_H^2}{\beta^2} \frac{r^2}{r_h^2}\, 
 {}_2F_1\left(1,\frac{D-2}{2},\frac{D+1}{2},\frac{r^2}{r_h^2}\right)
 \right]
 + \mathcal O(\beta^{-2}) \ . 
\end{equation}
The radial component of the metric behaves around the center $r=0$ as 
\begin{equation}
 g_{rr} 
 = 
 1 - \frac{4\,\Gamma^2\!\left(\frac{D-2}{2}\right) \beta^2}
 {(D-1)\pi\,\Gamma^2\!\left(\frac{D-3}{2}\right) r_h^2 \beta_H^2} r^2 + \cdots \ , 
\end{equation}
and is of $\mathcal O(1)$ for $r \ll \frac{\beta_H}{\beta} r_h$ 
but is suppressed as $\mathcal O(\beta^{-2})$ for $r\gg \frac{\beta_H}{\beta} r_h$. 
Thus, each of quantities which concerns the quantization condition has 
the same behavior to the case of $D=4$, in the order estimation. 
The quantization condition is roughly estimated 
in a similar fashion to the case of $D=4$ as 
\begin{equation}
 \frac{\beta_H^2 r_h}{2\pi \alpha' \beta} \sim 1 \ . 
\end{equation}
and hence, the size of the solution is approximately the same to the Schwarzschild radius, 
\begin{equation}
 r_0 \simeq r_h \sim \beta\ , 
\end{equation}
or equivalently, the temperature approaches the Hawking temperature in $\beta\to\infty$. 

Now, we calculate the ADM mass and entropy. 
By using \eqref{ADM}, the ADM mass is given by 
\begin{align}
 M 
 &= 
 \int d r \, r^{D-2} \, \Omega_{D-2} 
 \frac{\left(\beta_H^2 + (D-4) f(r)\beta^2\right) D_0}{4(D-3)\pi G_N \sqrt{h(r)}} 
 \notag\\
 &= 
 \frac{\beta_H^3 r_0^{D-1} D_0 \Omega_{D-2}}{8(D-1)(D-3) \pi G_N \beta \sqrt{h(r)}}
 \left(1 + \frac{D-4}{h_0}\right)
 \notag\\
 &\qquad\times
 \biggl[
 (D-2)\,{}_2F_1\left(\frac{1}{2},\frac{D-1}{2},\frac{D+1}{2},\xi(\beta)\right) 
 \notag\\
 &\qquad\qquad
 + (D-4)\left(1 - \xi(\beta)\right) {}_2F_1\left(\frac{3}{2},\frac{D-1}{2},\frac{D+1}{2},\xi(\beta)\right) 
 \biggr]
  \ . 
\end{align}
By taking $\xi(\beta)\to 1$, and substituting \eqref{h0-higher-low}--\eqref{r0-higher-low}, 
we obtain 
\begin{equation}
 M 
 = 
 \frac{(D-2) r_h^{D-3} \, \Omega_{D-2}}{16 \pi G_N} \ . 
\end{equation}
Thus, we have reproduced the ADM mass of the Schwarzschild spacetime as is expected. 
The entropy is calculated by using \eqref{Entropy} as 
\begin{align}
 S 
 &= 
 \frac{\beta^3}{4\pi G_N} 
 \int d r \, 
 r^{D-2} \, \Omega_{D-2} 
 f(r) h^{-1/2}(r) D_0 
 \notag\\
 &= 
 \frac{\beta_H^3 r_0^{D-1}}{4(D-1)\pi G_N\sqrt{h_0}} 
 \biggl[
  \frac{1}{h_0} {}_2F_1\left(\frac{3}{2},\frac{D-3}{2},\frac{D+1}{2},\xi(\beta)\right)
 \notag\\
 &\qquad\qquad\qquad\qquad\qquad 
  + \frac{\xi(\beta)}{D+1} {}_2F_1\left(\frac{3}{2},\frac{D-1}{2},\frac{D+3}{2},\xi(\beta)\right)
 \biggr]
  \ . 
\end{align}
By taking $\xi(\beta)\to 1$, and substituting \eqref{h0-higher-low}--\eqref{r0-higher-low}, 
we obtain 
\begin{equation}
 S = \frac{(D-3)\beta r_h^{D-3} \, \Omega_{D-2}}{16\pi G_N} \ . 
\end{equation}
By imposing the quantization condition, 
the temperature approaches the Hawking temperature in the low temperature limit. 
Then, we reproduce the Bekenstein-Hawking entropy up to the numerical coefficient, 
\begin{equation}
 S \sim \frac{r_h^{D-2} \, \Omega_{D-2}}{4 G_N} \ . 
\end{equation}


\section{Conclusion and discussions}\label{sec:Conclusion}

In this paper, we studied the transition between self-gravitating strings and black holes. 
It is expected that black holes turn into bound states of strings 
at a temperature near the string scale. 
Horowitz and Polchinski studied an effective field theory of winding strings 
which is wrapping the Euclidean time circle, and showed explicitly that there is 
a solution of such bound states slightly below the Hagedorn temperature. 
Near the Hagedorn temperature, strings are bounded weakly by the self-gravitation 
and the bound state can be described by the linearized gravity. 
However, gravitation becomes very strong around 
the transition between black holes and strings, 
and hence, non-linear effects of gravity should be taken into account. 

In order to study details of the black hole--string transition, 
we have proposed a fluid model of self-gravitating strings. 
Since the fluid is one of the most standard states with local equilibrium, 
it is expected that strings near the Hagedorn temperature behave as a fluid, at least approximately. 
In this paper, we first have shown that the stress-energy tensor 
of winding strings in the Horowitz-Polchinski model 
approximately takes the same form to perfect fluids. 
We have derived the stress-energy tensor in two different ways. 
First, we have shown that the stress-energy tensor of 
the ground state of the Horowitz-Polchinski model 
takes the form of the perfect fluid at the leading order of the derivative expansion. 
Alternatively, we also have calculated the energy density and pressure 
from the thermodynamic relations assuming that winding strings behave as a perfect fluid. 
The results of these two methods agree with each other. 
Thus, the stress-energy tensor of winding strings 
can be approximated by that of a perfect fluid. 
The fluid distribution extends only in the region where the pressure is positive. 
Since, the pressure of the fluid of winding strings becomes positive iff 
the local temperature exceeds the Hagedorn temperature due to the gravitational blue-shift. 
Thus, the fluid of winding strings localized in the region 
where the condensate of winding strings occurs. 

Next, we have considered the spacetime geometry of the fluid of winding strings. 
For simplicity, we have ignored the coupling to the dilaton field. 
Then, we have solved the Einstein equation for 
the stress-energy tensor of the winding string fluid. 
The interior solution is obtained analytically for arbitrary dimensions, 
and has no singularity by choosing the integration constant appropriately. 
We have taken the Schwarzschild spacetime as the exterior vacuum solution 
and have considered the junction conditions of these two solutions. 
The interface between these two solutions can be put 
in an arbitrary position but outside the Schwarzschild radius. 

In the solutions of the Horowitz-Polchinski model, 
physical configurations of strings would be 
described by the ground state of the winding string field, 
which satisfies the first quantization condition. 
In our fluid model, we do not consider the field equation of the winding string field 
and hence, the quantization condition is not imposed automatically. 
The quantization condition gives an additional constraint 
between the size of the star and the temperature of the fluid. 
In the high temperature limit, or equivalently near the Hagedorn temperature, 
the fluid has approximately the same size to the solution of the Horowitz-Polchinski model, 
by imposing the quantization condition. 
The ADM mass and entropy is also approximately the same to the Horowitz-Polchinski model. 

We also have considered the low temperature limit. 
Although the Horowitz-Polchinski model is expected to 
give a good description of strings only near the Hagedorn temperature, 
strings in bound states by the strong self-gravitation 
possibly have sufficiently high local temperatures due to the gravitational blue-shift, 
even if the global temperature is much lower than the Hagedorn temperature. 
In fact, our solution has a configuration whose 
local temperature exceeds the Hagedorn temperature, even in the low temperature limit. 
By imposing the quantization condition, 
the size of the star approaches the Schwarzschild radius in low temperature. 
Moreover the temperature and entropy of the fluid are also 
approximately the same to the Schwarzschild black hole. 
Although the geometry has no event horizon, 
the star of the winding string fluid approximately behaves as 
and is almost indistinguishable from a black hole from the viewpoint of 
fiducial observers sufficiently away from the star. 

As summarized above, we have constructed a solution of a star of winding strings. 
Our solution is given at arbitrary temperatures below the Hagedorn temperature, 
and varies smoothly under the change of the temperature. 
Winding strings are weakly bounded by self-gravitation in the high temperature limit, 
and approaches the Schwarzschild black hole in the low temperature limit. 
Thus, our solution gives a description of the transition between black holes and strings. 

It would be surprising that our solution is still valid at arbitrarily low temperatures 
though it is constructed as a bound state of winding strings. 
As there is the transition between self-gravitating strings and black holes, 
one would expect that models of winding strings would be valid 
only in the self-gravitating string phase and break down at the critical point. 
If this were the case, the black hole phase would be simply given by the Schwarzschild solution. 
In this sense, it would be reasonable that Horowitz and Polchinski focused 
only on the weakly bounded winding strings in the linearized gravity. 
If the Horowitz-Polchinski solution is naively extrapolated to 
the critical point between the self-gravitating string phase and black hole phase, 
the bound state of winding strings would fall inside the Schwarzschild radius, 
and hence, the winding string picture might be expected to break down around the critical point. 
However, our solution of the winding string fluid does not break down at the critical point. 

In this paper, we have taken non-linear effects of gravity into account 
and found that the size of the bound state is slightly larger than 
the Schwarzschild radius in the low temperature limit. 
Our solution is valid at arbitrarily low temperatures and 
gives a description even in the black hole phase, contrary to the naive expectation. 
The smoothness of the solution, as a function of the temperature, implies that the transition 
between the self-gravitating strings and black holes takes place continuously, 
and there is no discontinuity between two phases. 
In this sense, it might be inappropriate to call it a phase transition, 
but a discontinuity might be found by taking corrections into account. 

Although the horizonless geometry which approximately the same to the Schwarzschild black hole 
possibly turns into an exact black hole with the event horizon 
when the size of the star is indistinguishable from the Schwarzschild radius, 
it may alternatively indicate that real black holes have no exact event horizon. 
It is sometimes considered that such geometries cannot be realized 
because of the Buchdahl theorem \cite{Buchdahl:1959zz}. 
The Buchdahl theorem states that no static configuration of the perfect fluid
has the radius equal to or less than 9/8 of the Schwarzschild radius. 
However, our solution can violate the Buchdahl inequality. 
The Buchdahl theorem assumes that the energy density is non-increasing outward and non-negative. 
Our solution does not satisfy these assumptions. 
The assumption of the Buchdahl theorem is reasonable 
for fluids of ordinary matters but is not necessarily satisfied 
by strings near the Hagedorn temperature. 
The negative energy density comes from the condensate of winding strings, 
or equivalently, appears because winding strings become tachyonic when 
the temperature effectively exceeds the Hagedorn temperature. 
Effects of the winding string condensate are completely stringy 
and appear only near the string scale. 
However, these stringy effects possibly become important even if 
the Hawking temperature is much lower than the string (or Planck) scale 
because the local temperature is highly blue-shifted near black holes. 

In this paper, we have ignored the coupling with dilaton field. 
It would be interesting to see how the dilaton coupling will modify the results in this paper. 
Since the fluid of winding strings couples with the dilaton, 
the exterior solution would also be modified from the Schwarzschild spacetime. 
The fluid model itself is merely an approximation and 
winding strings do not behave as a perfect fluid exactly. 
It would also be interesting to see more precise behaviors of winding strings, 
though it is very difficult to solve the original equations of motion of the winding string field. 
Even in the fluid model, several issues are left to be studied in more detail. 
For example, the Horowitz-Polchinski model has no normalizable solution for $D\geq 7$, 
whereas we have not reproduced this non-normalizability in our fluid model. 
There are several models related to this problem including charged solutions. 
They are left for future studies.


\section*{Acknowledgments}

The author would like to thank Koji~Hashimoto and Takuya~Yoda 
for fruitful discussions on the Horowitz-Polchinski model.  
The author would also like to thank Akihiro~Ishibashi for valuable comments. 
This work is supported in part by JSPS KAKENHI Grant No.~JP17H06462, JP20K03930 and JP21H05186.

\end{document}